\documentclass[a4paper,11pt]{article}
\pdfoutput=1
\usepackage{jcappub}
\usepackage{fontawesome5}
\usepackage{mathtools}
\usepackage{xcolor}

\newcommand{\rR}{\rho_R}
\newcommand{\nbh}{n_\text{PBH}}
\newcommand{\Tbh}{T_\text{BH}}
\newcommand{\Mbh}{M_\text{PBH}}

\newcommand{\Tin}{T_\text{in}}

\DeclareMathOperator\erf{erf}
\definecolor{orcidlogocol}{rgb}{0.65, 0.807, 0.223}
\newcommand{\orcid}[1]{$\,$\href{https://orcid.org/#1}{\textcolor{orcidlogocol}{\faOrcid}}}

\begin{document}

\title{Current and future neutrino limits \\ on the abundance of \\ primordial black holes}

\author[a,\,b]{Nicolás Bernal\orcid{0000-0003-1069-490X},}
\author[c]{Víctor Muñoz-Albornoz\orcid{0000-0002-4142-2942},}
\author[c]{\\Sergio Palomares-Ruiz\orcid{0000-0001-9049-2288}}
\author[c,\,d]{and Pablo Villanueva-Domingo\orcid{0000-0002-0936-4279}}
\affiliation[a]{New York University Abu Dhabi\\
PO Box 129188, Saadiyat Island, Abu Dhabi, United Arab Emirates}
\affiliation[b]{Centro de Investigaciones, Universidad Antonio Nariño\\
Carrera 3 Este \# 47A-15, Bogotá, Colombia}
\affiliation[c]{Instituto de F\'isica Corpuscular (IFIC), Universitat de Val\`encia -- CSIC, \\ Parc Científic UV, C/ Catedrático José Beltrán, 2, E-46980 Paterna, Spain}
\affiliation[d]{Computer Vision Center (CVC) - Universitat Aut\`onoma de Barcelona (UAB),\\ 08193 Bellaterra (Barcelona), Spain}

\emailAdd{nicolas.bernal@nyu.edu}
\emailAdd{victor.manuel.munoz@ific.uv.es}
\emailAdd{sergiopr@ific.uv.es}
\emailAdd{pablo.villanueva.domingo@gmail.com}

\abstract{
Primordial black holes (PBHs) formed in the early Universe are sources of neutrinos emitted via Hawking radiation. Such astrophysical neutrinos could be detected at Earth and constraints on the abundance of comet-mass PBHs could be derived from the null observation of this neutrino flux. Here, we consider non-rotating PBHs and improve constraints using Super-Kamiokande neutrino data, as well as we perform forecasts for next-generation neutrino (Hyper-Kamiokande, JUNO, DUNE) and dark matter (DARWIN, ARGO) detectors, which we compare. For PBHs less massive than $\sim \textrm{few} \times 10^{14}$~g, PBHs would have already evaporated by now, whereas more massive PBHs would still be present and would constitute a fraction of the dark matter of the Universe. We consider monochromatic and extended (log-normal) mass distributions, and a PBH mass range spanning from $10^{12}$~g to $\sim 10^{16}$~g.  Finally, we also compare our results with previous ones in the literature.} 

\begin{flushright}
    PI/UAN-2021-691FT \\
    IFIC/22-08
\end{flushright}

\maketitle

\section{Introduction} 
\label{sec:intro}

The recent advances in the field of gravitational wave astronomy have renewed the interest from the scientific community in the possibility of discovering black holes (BHs) that would have been generated in the infancy of the Universe, before big bang nucleosynthesis (BBN). These primordial black holes (PBHs), first considered decades ago~\cite{Zeldovich:1967lct, Hawking:1971ei}, have gathered even more attention after the first detection by LIGO and VIRGO of gravitational waves emitted by the merger of a binary BH system~\cite{Abbott:2016blz}. The question whether the observed merger events correspond to BHs of astrophysical or primordial origin is still under debate, and a mixed population may also be compatible with observations~\cite{DeLuca:2020qqa, Hall:2020daa, Garcia-Bellido:2020pwq, Wong:2020yig, Hutsi:2020sol, DeLuca:2021wjr}.

PBHs present several observational effects that makes them fascinating objects, and allow us to place stringent bounds on their abundance for a broad range of masses, from ton-mass PBHs to super-massive BHs. Among the myriad of probes employed to bound the allowed population of PBHs, one can highlight the non-observation of microlensing events~\cite{Alcock:2000kg, Tisserand:2006zx, Griest:2013aaa, Oguri:2017ock, Niikura:2019kqi, Croon:2020ouk}, accretion signatures on the cosmic microwave background (CMB)~\cite{Poulin:2017bwe, Serpico:2020ehh} or on the EDGES 21-cm signal~\cite{Hektor:2018qqw}, disruption of stellar systems by the presence of PBHs~\cite{Monroy-Rodriguez:2014ula, Brandt:2016aco}, effects on the Lyman-$\alpha$ forest~\cite{Murgia:2019duy}, and merger rates from gravitational waves~\cite{Raidal:2017mfl, Ali-Haimoud:2017rtz, Kavanagh:2018ggo, Raidal:2018bbj, Chen:2019irf, Authors:2019qbw, Vaskonen:2019jpv, Hutsi:2020sol}. Moreover, future experiments could improve these bounds, as different forecasts show for next generation gravitational waves detectors~\cite{Guo:2017njn, Jung:2017flg, Bartolo:2018rku, Pujolas:2021yaw, Urrutia:2021qak}, spectral distortions of the CMB~\cite{Nakama:2017xvq} and 21-cm cosmology~\cite{Mena:2019nhm, 2022PASJ..tmp....9V} (see, e.g., Refs.~\cite{Sasaki:2018dmp, Green:2020jor, Carr:2020gox, Carr:2020xqk, Villanueva-Domingo:2021spv} for recent reviews regarding current constraints, formation and abundance of PBHs). 

A particularly captivating effect is the evaporation of BHs. Due to quantum effects in curved spacetime, BHs emit particles at their event horizon, which is known as Hawking radiation~\cite{Hawking:1974rv, Hawking:1974sw}. PBH evaporation due to Hawking radiation allows setting stringent bounds on their abundance for PBH masses $\Mbh \lesssim 10^{17}$~g, mainly through the observation of the extragalactic X-ray and $\gamma$-ray background~\cite{Carr:2009jm, Ballesteros:2019exr, Arbey:2019vqx, Iguaz:2021irx, Chen:2021ngo, Mosbech:2022lfg}, radio signals~\cite{Dutta:2020lqc}, their impact on BBN abundances~\cite{Kohri:1999ex, Carr:2009jm, Acharya:2020jbv}, anisotropies and spectral distortions of the CMB spectrum~\cite{Poulin:2016anj, Clark:2016nst, Stocker:2018avm, Poulter:2019ooo, Lucca:2019rxf, Acharya:2020jbv}, INTEGRAL and COMPTEL observations of the Galactic Center~\cite{DeRocco:2019fjq, Laha:2019ssq, Dasgupta:2019cae, Laha:2020ivk, Coogan:2020tuf}, electron and positron detections at Voyager~1~\cite{Boudaud:2018hqb}, and the claimed EDGES 21-cm signal~\cite{Clark:2018ghm, Halder:2021jiv, Halder:2021rbq, Mittal:2021egv, Natwariya:2021xki, Cang:2021owu, Saha:2021pqf}.

Neutrinos would also be emitted as Hawking radiation during the evaporation of PBHs and henceforth, could be detected at neutrino observatories. Neutrino astronomy was initiated with the simultaneous detection of $\sim 25$ neutrino events originated from the supernova 1987A at the Kamiokande~\cite{Hirata:1987hu}, IMB~\cite{VanDerVelde:1987hh} and Baksan~\cite{Alekseev:1988gp} detectors. Since then, many efforts have been devoted to seek for more neutrinos emitted at supernova occurrences~\cite{Mirizzi:2015eza, Horiuchi:2017sku}, from the diffuse supernova background (DSNB)~\cite{Beacom:2010kk, Lunardini:2010ab}, and from other astrophysical sources~\cite{Meszaros:2017fcs, Halzen:2022pez}. Future neutrino detectors, like Hyper-Kamiokande~\cite{Abe:2018uyc}, JUNO~\cite{Djurcic:2015vqa} and DUNE~\cite{Abi:2020wmh} may become indispensable for detecting low-energy (in the range of tens of MeV) neutrinos of cosmic origin. Moreover, detectors for direct dark matter (DM) detection, such as DARWIN~\cite{Aalbers:2016jon} and ARGO~\cite{Agnes:2020pbw}, may reach a good enough sensitivity to measure the neutrino floor and detect neutrino events from cosmic origin and others via coherent elastic neutrino-nucleus scattering (CE$\nu$NS)~\cite{McKeen:2018pbb, Gelmini:2018gqa, Raj:2019wpy, Raj:2019sci, Munoz:2021sad, Suliga:2021hek}.

Indeed, the detection of neutrino events has been proposed to probe the existence of comet-mass PBHs and to set limits on their abundance~\cite{Halzen:1995hu, Bugaev:2000bz, Bugaev:2001hd, Bugaev:2002yk, Bugaev:2002yt, Bugaev:2004yp}. Constraints on the PBH abundance with the same data used to search for the DSNB at Super-Kamiokande (SK)~\cite{Super-Kamiokande:2002hei, Bays:2011si, Super-Kamiokande:2021jaq} were obtained more than a decade ago~\cite{Carr:2009jm} and have been recently reviewed (including rotating PBHs)~\cite{Dasgupta:2019cae} using the maximum allowed integrated fluxes of the DSNB with current SK data~\cite{Bays:2011si}. The prospects of detecting neutrinos from PBHs at future neutrino detectors such as JUNO~\cite{Wang:2020uvi}, DUNE and THEIA~\cite{DeRomeri:2021xgy}, IceCube~\cite{Halzen:1995hu, Capanema:2021hnm} and at forecoming DM detectors such as DARWIN~\cite{Calabrese:2021zfq}, have also been recently studied. In this article, we revisit and improve upon the current bounds on the PBH abundance using SK data. We perform a complete spectral analysis, in analogy to one of the SK studies of the DSNB~\cite{Bays:2011si, Bays:2012wty} and update constraints on the fraction of PBHs that would constitute part of the DM today and on the abundance at production of those already evaporated. Moreover, we provide a comprehensive study of the sensitivities that can be reached by the most relevant future neutrino and DM detectors to constrain the abundance of PBHs, and extend the study to masses corresponding to PBHs with a lifetime shorter than the age of the Universe. For all detectors and PBH masses, we present results for monochromatic and log-normal mass distributions.

The structure of the article is as follows. Fundamentals on the formation and evaporation of PBHs are briefly discussed in Sec.~\ref{sec:PBH}, while the emitted spectrum and flux of neutrinos from PBH evaporation are discussed in Sec.~\ref{sec:neutrinoflux}. Then, in Sec.~\ref{sec:SK} we update and extend PBH bounds from SK data using the full spectral information of both, signal and backgrounds. The expected neutrino rates at forthcoming neutrino (Hyper-Kamiokande, JUNO, DUNE) and DM (DARWIN, ARGO) detectors are discussed in Sec.~\ref{sec:rates}, while the analyses of the prospects of detection with these future detectors and the forecast of constraints on the PBH abundance in the mass range $\Mbh \simeq (10^{12} - 10^{16})$~g is conducted in Sec.~\ref{sec:sensitivity}. Finally, a summary and concluding remarks are presented in Sec.~\ref{sec:conclusions}.
\pagebreak

\section{Evaporation of primordial black holes}
\label{sec:PBH}

If PBHs exist and they formed in a radiation dominated epoch, when the Standard Model (SM) plasma had a temperature $T = \Tin$, their initial mass would have been similar to the enclosed mass in the particle horizon ($M_{\rm H}$)~\cite{Carr:1975qj},\footnote{A more accurate treatment should account for the fact that the PBH mass at formation also depends on the value and shape of the collapsing overdensity~\cite{Choptuik:1992jv, Evans:1994pj, Niemeyer:1997mt, Niemeyer:1999ak, Musco:2004ak}.}
\begin{equation}
    \Mbh = \gamma \, M_{\rm H} = \frac{4\pi}{3}\, \gamma\, \frac{\rR(\Tin)}{H^3(\Tin)} ~,
\end{equation}
where $\gamma \simeq c_s^{3/2} \simeq 0.2$, with $c_s \simeq 1/3$ the speed of sound at the radiation epoch, is a numerical factor that depends on the details of the gravitational collapse, $\rR(\Tin)$ corresponds to the SM radiation energy density at the formation time, and $H$ to the Hubble expansion rate. PBHs can gain mass via mergers~\cite{Zagorac:2019ekv, Hooper:2019gtx, Hooper:2020evu} and accretion~\cite{Bondi:1952ni, Nayak:2009wk, Masina:2020xhk}, although these processes are typically not very efficient, only inducing a mass gain by a factor of order $\mathcal{O}(1)$, and will be, hereafter, ignored. At their formation time (when the plasma temperature is $\Tin$), the fraction of the Universe's energy density in the form of PBHs is~\cite{Carr:2009jm}
\begin{equation}
    \beta \equiv \frac{\rho_{\rm PBH}(\Tin)}{\rho_R(\Tin) + \rho_{\rm PBH}(\Tin)} \simeq \frac{\rho_{\rm PBH}(\Tin)}{\rho_R(\Tin)} ~.
\end{equation}
Generalizing the expression from Ref.~\cite{Carr:2009jm} for a broad mass spectrum, this is given by
\begin{align}
    \beta & \equiv n_{\rm PBH}(t_0)\int dM\, \frac{M}{\rho_R(T_{\rm in})} \, \frac{d\cal N}{dM} \nonumber \\
    & \simeq 2.86 \times 10^{-56}\, \gamma^{-1/2}\left(\frac{g_\star(\Tin)}{106.75} \right)^{1/4} \, \left(\frac{n_{\rm PBH}(t_0)}{{\rm Gpc}^{-3}}\right) \, \int dM \left(\frac{M}{10^{15}~{\rm g}} \right)^{3/2} \, \frac{d\mathcal{N}}{dM} ~,
\end{align}
where $g_\star(\Tin)$ counts the number of relativistic degrees of freedom contributing to the SM energy density at production time~\cite{Drees:2015exa}, which is different for every mass. At production time, the mass distribution function is $d\mathcal{N}/dM$, and $n_{\rm PBH}(t_0)$ is the PBH comoving number density today, had those PBHs not evaporated away. We consider the mass distribution function to be normalized to one, so that $\rho_{\rm PBH} = n_{\rm PBH} \int dM M \, d{\cal N}/dM = n_{\rm PBH} \, \overline{M}$, with $\overline{M}$ the mean of the distribution.

As the PBHs energy density scales as non-relativistic matter, if massive enough (i.e., if not fully evaporated, $M > M_{\rm evap}$, with $M_{\rm evap}$ the mass of PBHs evaporating now, see below), they would constitute part of the total DM abundance today, representing a fraction
\begin{equation}
f_{\rm PBH} \equiv \frac{\Omega_{\rm PBH}}{\Omega_{\rm DM}} \simeq \left( \frac{\beta'}{1.9 \times 10^{-18}} \right) \, \left( \frac{0.27}{\Omega_{\rm DM}} \right) \, \frac{\int_{M_{\rm evap}} dM \, \left(\frac{M}{10^{15}~{\rm g}} \right) \, \frac{d\mathcal{N}}{dM}}{\int d M \, \left(\frac{M}{10^{15}~{\rm g}} \right)^{3/2} \, \frac{d\mathcal{N}}{dM}} ~,
\end{equation}
and~\cite{Carr:2020gox}
\begin{equation} 
\label{eq:betaprime}
\beta' \equiv \gamma^{1/2}\left(\frac{g_\star(\Tin)}{106.75} \right)^{-1/4} \, \left( \frac{h}{0.67} \right)^{-2} \, \beta ~,
\end{equation}
where $h$ is the reduced Hubble constant. Note that, by setting the lower limit of the integral in the numerator to $M_{\rm evap}$, we only count (approximately) non-evaporated PBHs as part of the DM density today.

Note that, although monochromatic mass distributions are typically assumed to set limits on the PBH abundance, in general, PBH masses are expected to span over a broad interval, and constraints could be strengthened~\cite{Clesse:2015wea, Carr:2016drx, Green:2016xgy, Kuhnel:2017pwq, Carr:2017jsz, Bellomo:2017zsr, Carr:2020gox}. Extended PBH mass functions arise naturally if PBHs are created from inflationary fluctuations or cosmological phase transitions (see, e.g., Refs.~\cite{Hawking:1982ga, Carr:1994ar, Garcia-Bellido:1996mdl, Yokoyama:1998xd, Niemeyer:1999ak, Drees:2011hb, Kohri:2012yw, Kuhnel:2015vtw, Deng:2017uwc}), and a log-normal distribution, first obtained in Ref.~\cite{Dolgov:1992pu}, represents a very good approximation for models with a symmetric peak in the power spectrum of curvature perturbations~\cite{Green:2016xgy, Kannike:2017bxn}. In the analysis presented in the following sections, both monochromatic and log-normal distributions are considered, with
\begin{equation}
    \frac{d\mathcal{N}}{dM} =
    \begin{dcases}
        \delta\left(M - \Mbh\right)  & {\rm (monochromatic)} ~, \\[1.5ex]
        \frac{1}{\sqrt{2\pi}\, \sigma\, M} \exp\left[-\frac{\ln^2\left(M/\Mbh\right)}{2\, \sigma^2} \right]  & \textrm{(log-normal)} ~,
    \end{dcases}
\end{equation}
where $\Mbh$ denotes the median mass and $\sigma$ represents the width of the log-normal distribution.\footnote{Note that log-normal distributions for the comoving mass density $M d \mathcal{N}/dM$, rather than for the comoving number density, have been also considered (e.g., Refs.~\cite{Green:2016xgy, Kuhnel:2017pwq}).} The spectrum is normalized such that $\int dM d\mathcal{N}/dM = 1$. 

Therefore, the initial abundance $\beta'$ can be explicitly written in each case as
\begin{equation}
    \beta' \simeq 2.86 \times 10^{-56} \left( \frac{h}{0.67}\right)^{-2} \left(\frac{n_{\rm PBH}(t_0)}{{\rm Gpc}^{-3}}\right) \left(\frac{\Mbh}{10^{15}~{\rm g}} \right)^{3/2} \times \begin{dcases} \, 1 & {\rm (monochromatic)} , \\[1.5ex]
    \, e^{9 \,\sigma^2/8} & \textrm{(log-normal)} ,
     \end{dcases}
\end{equation}
and the relation between $f_{\rm PBH}$ and $\beta'$, as
\begin{equation}
\label{eq:fPBH}
f_{\rm PBH} \simeq 
\left( \frac{\beta'}{1.9 \times 10^{-18}} \right) \, \left( \frac{0.27}{\Omega_{\rm DM}} \right) \, \left(\frac{10^{15}~{\rm g}}{M_{\rm PBH}}\right)^{1/2} \times
\begin{dcases} \, 1 & {\rm (monochromatic)} ~, \\[1.5ex]
\, e^{-5 \,\sigma^2/8} \, {\cal S}_{\rm PBH}(\sigma) & \textrm{(log-normal)} ~,
\end{dcases}
\end{equation}
where the term
\begin{equation}
\label{eq:SPBH}
{\cal S}_{\rm PBH}(\sigma) = \frac{1}{2} \, \left(1 + \erf\left[\frac{\sigma^2 - \ln\left(\frac{M_{\rm evap}}{\Mbh}\right)}{\sqrt{2} \, \sigma}\right] \right)
\end{equation}
represents the fraction of PBHs that would not have evaporated by now (assuming instant evaporation).

Evaporation occurs because, due to quantum effects in curved spacetime, PBHs emit particles at their event horizon, the so-called Hawking radiation~\cite{Hawking:1974rv, Hawking:1974sw}. The emitted spectrum has a nearly thermal black-body spectrum, with a temperature $\Tbh$ given by
\begin{equation}
\label{eq:temperature}
\Tbh = \frac{M_P^2}{\Mbh} \simeq 10~\text{MeV} \left(\frac{10^{15}~\text{g}}{\Mbh}\right) ,
\end{equation}
where $M_P = (8\pi G_N)^{-1/2}$ is the reduced Planck mass, with $G_N$ the gravitational constant. The number $N_i$ of particles $i$ with $g_i$ degrees of freedom radiated per unit of time and energy is given by a quasi-thermal black-body spectrum~\cite{Hawking:1974rv, Hawking:1974sw},
\begin{equation} 
    \left.\frac{d^2N_i(E,t)}{dt\,dE}\right|_{\rm prim} = \frac{g_i}{2\pi} \frac{\Gamma_i(E,\Mbh) }{e^{E/\Tbh} \pm 1} ~,
\end{equation}
with $-1$ and $+1$ for bosons and fermions, respectively, and $\Gamma_i$ is a gray-body factor~\cite{Teukolsky:1974yv, Page:1976df, Page:1976ki, Page:1977um}. In addition to this primary spectrum generated via Hawking evaporation, the decay of hadrons and other particles promptly produced, results in a secondary spectrum of stable particles, including electrons, neutrinos and photons.

The neutrino spectrum rate at time $t$ is given by the sum of these two contributions weighted by the PBH mass distribution,
\begin{equation} \label{eq:dNdEdt}
    \frac{d^2N_\nu}{dE_\nu\, dt} = \int_0^\infty dM \, \frac{d\mathcal{N}}{dM} \left( \frac{d^2N_\nu}{dE_\nu\, dt}{\Big|}_{\rm prim} + \frac{d^2N_\nu}{dE_\nu\, dt}{\Big|}_{\rm sec} \right)  ~.
\end{equation}
We compute the neutrino emission rates via the publicly available code {\tt BlackHawk}~\cite{Arbey:2019mbc, Auffinger:2020ztk, Arbey:2021mbl}. To obtain the secondary spectra resulting from hadronization, interconversion, and decays of particles from the primary emission, {\tt BlackHawk} makes use of {\tt PYTHIA}~\cite{Sjostrand:2014zea}, {\tt HERWIG}~\cite{Bellm:2015jjp} and {\tt Hazma}~\cite{Coogan:2019qpu}. We consider both contributions, which are computed for each PBH mass with {\tt BlackHawk v1.2}.\footnote{\label{fn:BHoutput} Note that the output of {\tt BlackHawk} for the secondary spectra already includes the primary ones (private communication with {\tt BlackHawk}'s authors).} Notice that, whereas in {\tt BlackHawk v1.2} the degrees of freedom of light quarks are accounted for by considering the emission of single quarks (with free quark masses) below the QCD scale, {\tt BlackHawk v2.1}~\cite{Arbey:2021mbl} enables the choice of a new low-energy hadronization scheme in which pions instead of quarks are emitted below the QCD scale. In principle, the approach in {\tt BlackHawk v1.2} could overestimate the number of degrees of freedom for masses $10^{14}~{\rm g} \lesssim M_{\rm PBH} \lesssim 10^{16}$~g (see, e.g., Ref.~\cite{Stocker:2018avm}), with respect to the approach which considers effective quark and gluon masses and only considers pion emission below the QCD scale~\cite{MacGibbon:1990zk, MacGibbon:1991tj, MacGibbon:2007yq}. This is expected to result in different neutrino fluxes (see Refs.~\cite{Coogan:2020tuf, Arbey:2021mbl} for the comparison of secondary photon and electron emission between both approaches) and therefore, in bounds which differ by a non-negligible amount from the ones we present here (or that have been obtained in other works with the use of {\tt BlackHawk v1.2}, in the mentioned mass range). Nevertheless, {\tt BlackHawk v2.1} does not provide low-energy neutrino spectra using the new approach, yet. Therefore, with this note of caution, we proceed and show results obtained using the neutrino fluxes computed with {\tt BlackHawk v1.2}, assuming Majorana neutrinos (six degrees of freedom). Note that, although the detection events are not sensitive to their Dirac/Majorana nature, in the case of Dirac neutrinos, right-handed and left-handed neutrinos are emitted in equal amounts (twelve rather than six degrees of freedom)~\cite{Lunardini:2019zob}, which results in slightly higher fluxes.

\begin{figure}
	\centering
	\includegraphics[scale=0.55]{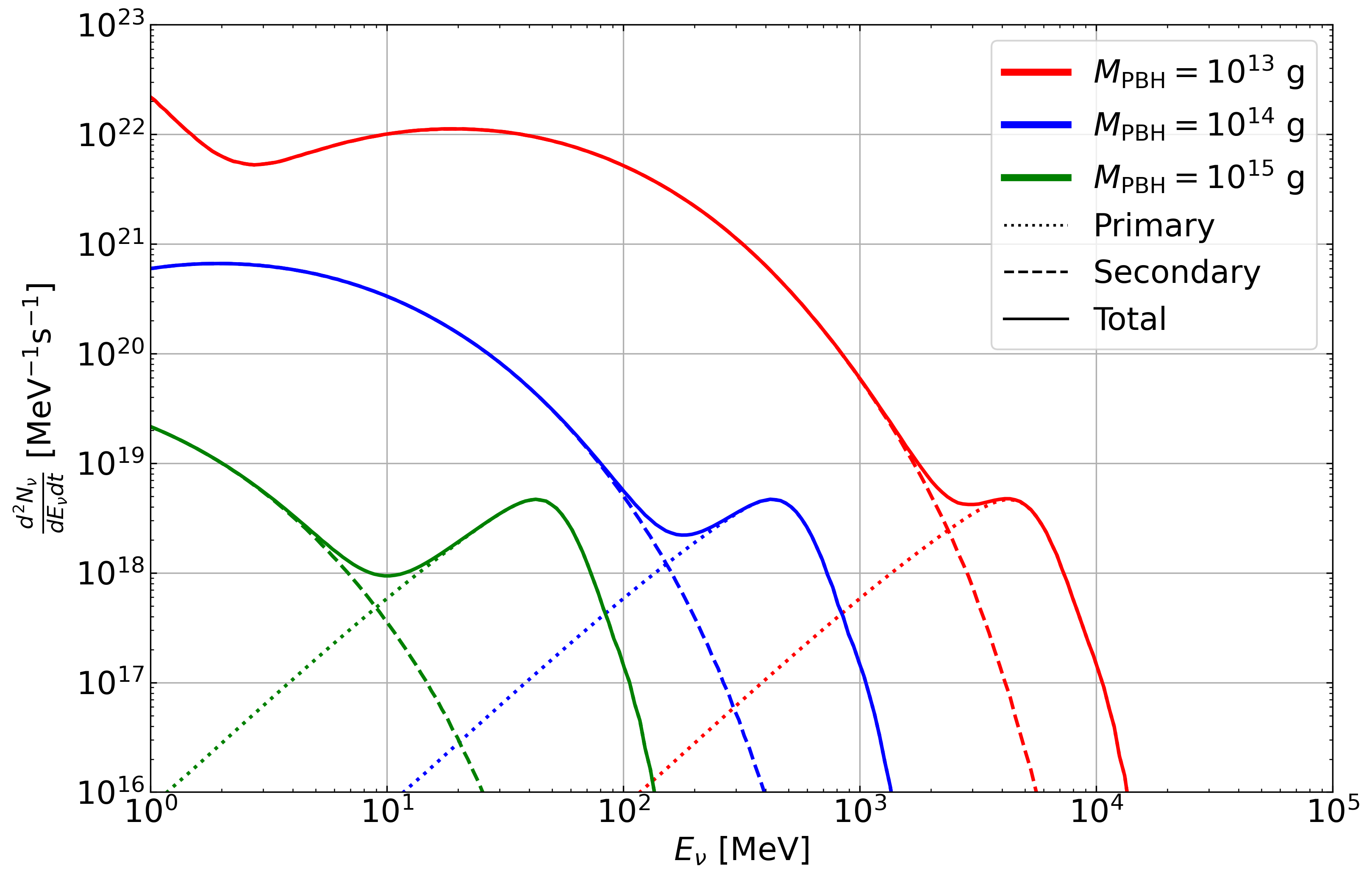}
	\caption{Electron neutrino emission rate from PBH evaporation, for three PBH masses, as a function of energy, explicitly showing the primary (dotted lines) and secondary (dashed lines) spectra, as well as their sum (solid lines).}
	\label{fig:emis}
\end{figure}

Figure~\ref{fig:emis} depicts the emission rate of $\nu_e$ for several PBH masses, explicitly showing the primary and secondary contributions, along with the total one. The primary spectrum dominates at high energies, around its peak at $E \sim 4.5 \, \Tbh$~\cite{Page:1976df}, approaching the geometric optics limit, $g_i \, \Gamma_i \simeq 27\, G_N^2\, M_{\rm PBH}^2\, E^2$, beyond the peak~\cite{Page:1976df, MacGibbon:1990zk}. However, secondary spectra dominate at lower energies, and become more prominent the smaller the PBH mass. Nonetheless, as discussed below, for most masses, the impact of the secondary spectrum on the results is mild, since backgrounds dominate the potential signal at energies where secondary spectra lie.

Due to this emission, PBHs continuously lose mass until their complete evaporation, with a lifetime that is approximately given by~\cite{MacGibbon:1991tj, MacGibbon:2007yq}
\begin{equation}
\label{eq:lifetime}
\tau \simeq 8.2 \, \tau_{\rm U} \, \left(\frac{M_{\rm PBH}}{10^{15}~{\rm g}}\right)^3,
\end{equation}
where $\tau_{\rm U} \simeq 13.8$~Gyr is the current age of the Universe. Hence, from the above estimate, PBHs with masses around $\sim 5 \times 10^{14}$~g would have a lifetime that equals the current age of the Universe. Note, however, that the treatment of the number of degrees of freedom just below the QCD scale in the {\tt BlackHawk v1.2} code leads to a mass of PBH evaporating now of $M_{\rm evap} \simeq 8 \times 10^{14}$~g. If their initial mass is $\Mbh \lesssim M_{\rm evap} $, their lifetime is smaller than the age of the Universe, and therefore they would have already (or almost) fully evaporated. Alternatively, for initial masses $\Mbh \gtrsim M_{\rm evap}$, PBHs would have not evaporated yet. In such a case, PBHs would unavoidably form part of the DM today.

\section{Neutrino flux from PBH evaporation}
\label{sec:neutrinoflux}

Given the emission rate $d^2N_\nu/(dE_\nu dt)$ from Eq.~\eqref{eq:dNdEdt}, the differential neutrino flux that reaches Earth is computed by including two contributions: the spectrum from the evaporation of extragalactic PBHs, taking into account the redshifting of energy, $\Phi_{\rm EG}(E_\nu)$, and the spectrum of neutrinos from non-evaporated PBHs within the Milky Way, $\Phi_{\rm MW}(E_\nu)$. The total flux is given by the sum of both contributions, $\Phi(E_\nu) = \Phi_{\rm EG}(E_\nu) + \Phi_{\rm MW}(E_\nu)$. The differential extragalactic flux includes the contribution over the full sky integrated in redshift, and reads~\cite{Carr:2009jm, Arbey:2019vqx},\footnote{This expression comes from the solution of the evolution equation accounting for redshift, which in terms of the intensity $I = E \, \frac{d\Phi_{\rm EG}}{dE_\nu}(E_\nu)$, takes the form $\frac{\partial I}{\partial t} -HE\frac{\partial I}{\partial E} = E \,\nbh(t_0)\frac{d^2N_\nu}{dE_\nu dt} $.}
\begin{equation}
\frac{d\Phi_{\rm EG}(E_\nu)}{dE_\nu} = \nbh(t_0) \int_{z_{\rm max}}^{z_{\rm min}} dz \left|\frac{dt}{dz}\right| (1+z) \, \frac{d^2N_\nu \left(E_\nu \, (1+z), \, t\right)}{dE_\nu dt} ~, 
\label{eq:flux}
\end{equation}
where $z_{\rm min}$ is the redshift corresponding to either today ($z_{\rm min} = 0$) or to the latest time in the output ($d^2N_\nu/dE_\nu dt$) from {\tt BlackHawk} in case PBHs would have already evaporated away (i.e., $z_{\rm min} > 0$), and $z_{\rm max}$ corresponds to the redshift at which PBHs where formed.
Note, however, that neutrinos from PBH evaporation emitted before neutrino decoupling ($t \sim 1$~s) do not contribute to this flux. The precise initial time used in the above integral is not important because of the redshifting of energy.

As already discussed, for $M_{\rm PBH} > M_{\rm evap}$, PBHs would not have fully evaporated yet, and would contribute to the DM density now. For those masses, the current PBH number density can be written as a function of the DM fraction contributed by PBHs now, as ${\cal S}_{\rm PBH}(\sigma) \, \nbh(t_0) = f_{\rm PBH} \, \rho_{\rm DM}/\overline{M}$, where $\overline{M}$ is the mean of the mass distribution and ${\cal S}_{\rm PBH}(\sigma)$ is given by Eq.~(\ref{eq:SPBH}): $\overline{M} = \Mbh$ (${\cal S}_{\rm PBH} = 1$) for the monochromatic case and $\overline{M} = \Mbh \, {\rm exp}(\sigma^2/2)$ for log-normal mass distributions. For those masses, PBHs in the Milky Way would also contribute to the neutrino emission, with a differential flux given by
\begin{equation}
\frac{d\Phi_{\rm MW}(E_\nu)}{dE_\nu} = \frac{d^2N_\nu(E_\nu,t_0)}{dE_\nu dt} \, \frac{f_{\rm PBH}}{\overline{M}} \int \frac{d\Omega}{4\pi} \int_0^{\ell_{\rm max}} d\ell\, \rho_{\rm NFW}\left(r(\ell,\phi)\right) ~,
\end{equation}
where $d^2N_\nu(E_\nu, t_0)/dE_\nu dt$ is the {\tt BlackHawk} output for the instantaneous neutrino spectrum at $z = 0$, $r(\ell,\phi)=\sqrt{R_\odot^2 + \ell^2 - 2 \, r_\odot \, \ell \, \cos\phi}$ is the radial coordinate, $R_\odot \simeq 8.5$~kpc the distance from the center of the Milky Way to the solar system,\footnote{Note that this (standard) value is slightly larger than the current best estimates~\cite{GRAVITY:2021, Bobylev:2021}.} $\ell$ the line-of-sight distance, and $\phi$ the angle between these two directions. Note that the double integral is known as the D-factor in decaying DM literature (see, e.g., Ref.~\cite{Evans:2016xwx}). The line-of-sight integral extends up to $\ell_{\rm max} = r_\odot \, \cos\phi + \sqrt{R_h^2 - R_\odot^2 \, \sin^2\phi} $, with $R_h \simeq 200$~kpc being the halo radius. In what follows, the DM density profile is approximated by the Navarro-Frenk-White (NFW) profile~\cite{Navarro:1995iw, Navarro:1996gj},
\begin{equation}
\rho_{\rm NFW}(r) = \rho_\odot \, \left( \frac{r_\odot}{r} \right) \, \left(\frac{1+R_\odot/r_s}{1 + r/r_s}\right)^2.
\end{equation}
For the sake of comparison with other works in the literature, we use $\rho_\odot = 0.4 \, {\rm GeV/cm}^3$ for the DM density in the solar neighborhood, and $r_s = 20$~kpc for the scale radius of the Milky Way.\footnote{Note, however, that this combination of values is at odds with current estimates~\cite{Benito:2020lgu, deSalas:2020hbh}. Furthermore, they are correlated with the galactocentric distance $R_\odot$~\cite{Benito:2019ngh}, which we assume to be slightly larger than present estimates~\cite{GRAVITY:2021, Bobylev:2021}. From this point of view, our results are conservative.} This contribution can be of the same order or larger than the extragalactic one, and thus cannot be neglected, in general, for non-evaporated PBHs.

Furthermore, note that neutrinos of a given flavor produced via PBH evaporation would mix with the other flavors along the way to Earth via neutrino oscillations. Denoting by $P_{\rm avg}(\nu_\beta \rightarrow \nu_\alpha) = \sum_i |U_{\alpha, i}|^2 \, |U_{\beta, i}|^2$ the probability of averaged oscillations between flavors $\nu_\beta$ and $\nu_\alpha$ (with $U$ the neutrino mixing matrix), the differential flux of neutrinos of flavor $\alpha$, $d\Phi_{\nu_\alpha}/dE_\nu$, is given by the combination of the non-oscillated neutrino fluxes,
\begin{equation}
\frac{d\Phi_{\nu_\alpha}(E_\nu)}{dE_\nu} = \sum_\beta P_{\rm avg}(\nu_\beta \rightarrow \nu_\alpha) \, \left. \frac{d\Phi_{\nu_\beta}(E_\nu)}{dE_\nu}\right|_{\rm no \, osc.},
\end{equation}
with neutrino mixing angles and CP phase taken from Ref.~\cite{deSalas:2020pgw} (see also Refs.~\cite{Esteban:2020cvm, Capozzi:2017ipn}). Nevertheless, the flux of secondary spectra at Earth, in the PBH mass range of interest, peaks at lower energies ($E_\nu \lesssim 10$ MeV) than the ones accessible by neutrino detectors, where the solar electron neutrino (and reactor electron antineutrino) background dominates. Thus, given that the primary spectrum for all neutrino flavors is the same, neutrino oscillations only modify neutrino fluxes at low energies (secondary spectrum) at the $\sim 2$\% level~\cite{Wang:2020uvi, DeRomeri:2021xgy}. For the range of masses we consider, this occurs at energies for which backgrounds dominate the signal and thus, oscillations have a negligible impact on the final results presented in this work.

\begin{figure}
	\centering
	\includegraphics[scale=0.51]{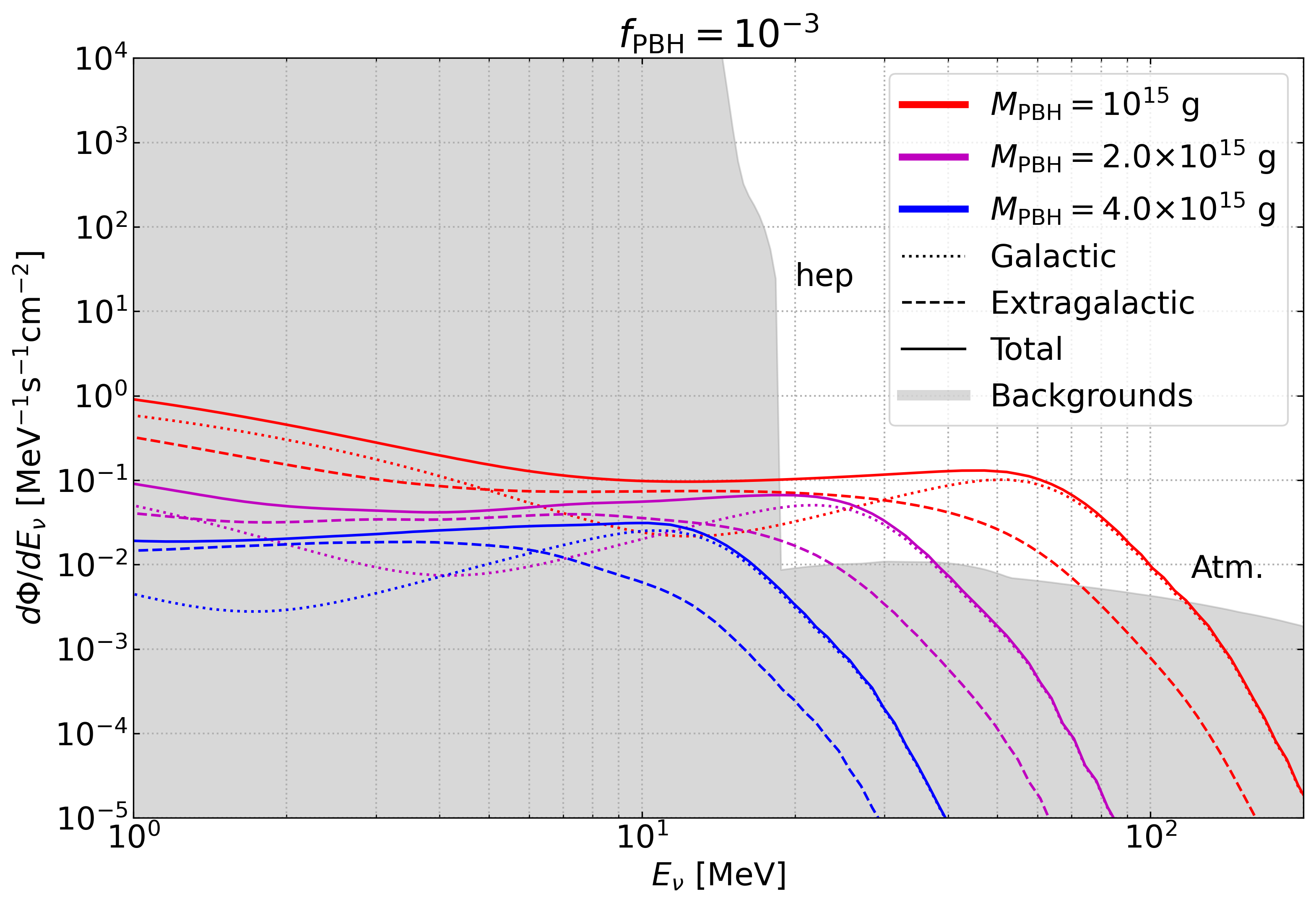}
	\includegraphics[scale=0.51]{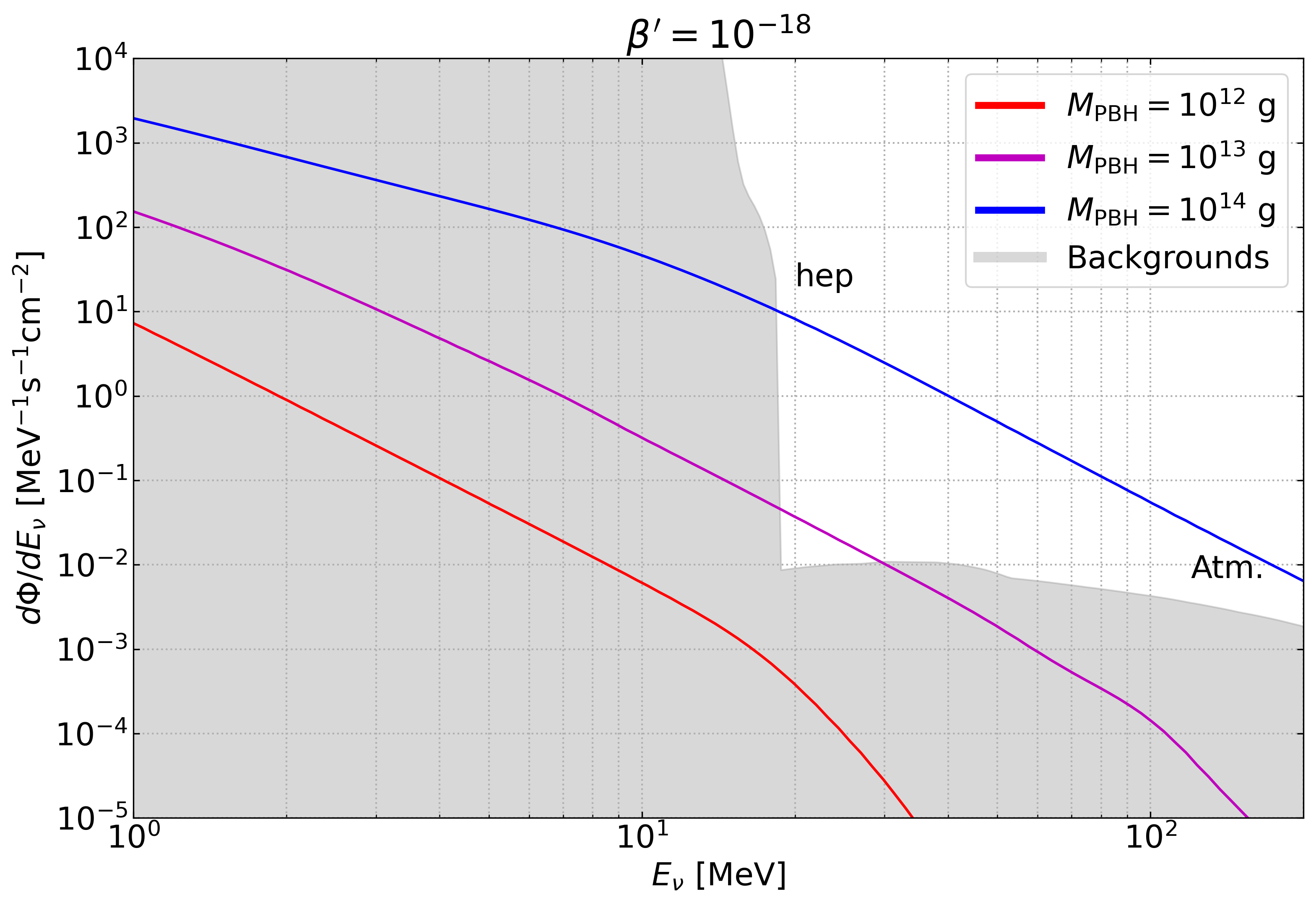}
	\caption{Electron neutrino fluxes from PBHs evaporation, as a function of energy, for non-evaporated PBHs with $f_{\rm PBH} = 10^{-3}$ (top panel) and for already evaporated PBHs with $\beta' = 10^{-18}$ (bottom panel), along with the most relevant solar and atmospheric electron neutrino fluxes (shaded regions). For non-evaporated PBHs, which would contribute to the DM density today, we show the galactic (dotted lines), extragalactic (dashed lines) and total (solid lines) contributions. For evaporated PBHs, only the extragalactic flux contributes. Both panels assume a monochromatic mass distribution. Note that, depending on the detection channels, the relevant backgrounds might be different from the ones shown here.}
	\label{fig:flux}
\end{figure} 

The electron neutrino flux from PBH evaporation is shown in Fig.~\ref{fig:flux}, where we separately depict the galactic, extragalactic and total contributions for PBHs as DM, $\Mbh > M_{\rm evap}$ (top panel) and the extragalactic flux for evaporated PBHs, $\Mbh < M_{\rm evap}$ (bottom panel). In both cases, the fluxes are shown for monochromatic mass distributions. For comparison, the solar neutrino flux, as predicted from the AGSS09 model~\cite{Robertson:2012ib}, and the atmospheric electron neutrino flux from the Monte-Carlo FLUKA simulations~\cite{Battistoni:2005pd}, are also shown. Another potential source of background for the PBH signal is the DSNB flux, which could be important for $E_\nu \sim 10 - 30$~MeV and whose impact is discussed below, but it is not included in the figure. 

The neutrino flux from non-evaporated PBHs ($\Mbh > M_{\rm evap}$) could dominate over other neutrino sources for $E_\nu \gtrsim 20$~MeV, whereas the PBH antineutrino flux (not shown) could be the highest flux for $E_\nu \gtrsim 10$~MeV. The galactic contribution peaks at higher energies than the extragalactic one, which is time-integrated and also experiences energy redshift. In the case of a monochromatic mass distribution, for $\Mbh \gtrsim 2 \times 10^{15}$~g, the galactic contribution would be the largest one for $E_\nu \gtrsim 15$~MeV, whereas for less massive PBHs, the extragalactic contribution could also be important at energies around $\sim 20$~MeV. On the other hand, the emission from PBHs already evaporated ($M_{\rm PBH} < M_{\rm evap}$), only has the extragalactic contribution, which peaks at even lower energies than that of more massive PBHs. At emission time, the initial spectra are warmer than for non-evaporated PBHs (given the lower PBHs masses). Nevertheless, after $z_{\rm evap}(M_{\rm PBH})$,  evaporation is almost complete and neutrinos are scarcely emitted. As can be understood from Eqs.~(\ref{eq:temperature}) and~(\ref{eq:lifetime}), the redshift in energy from $z > z_{\rm evap}(M_{\rm PBH})$ compensates the higher energies at emission, leading to a cooler final spectra (shifted to lower energies) at Earth, as shown in Fig.~\ref{fig:flux}, where only the high-energy tail of the spectrum reaches beyond $\sim 10$~MeV. 

In the case of the two extended mass distributions we consider in this work, the extragalactic contribution is always the dominant one at the energies of interest (i.e., a few tens of MeV), even in the case of median PBH masses $\Mbh > M_{\rm evap}$. Indeed, the more extended the mass distribution, the larger is the relative contribution from the extragalactic signal. This can be understood from the fact that the maximum of the mass distribution lies at lower masses the larger the width is. In this case, for median masses $\Mbh > M_{\rm evap}$, a fraction of PBHs in the low-mass tail would have evaporated by now and would not be part of the DM. This is (approximately) taken into account when setting limits on $f_{\rm PBH}$, see Eq.~(\ref{eq:fPBH}). Conversely, for median masses $\Mbh < M_{\rm evap}$, a fraction of PBHs in the high-mass tail of the distribution would have survived today and would constitute part of the DM now. Yet, in this case, we only quote limits on $\beta'$.

\section{Updated constraints with Super-Kamiokande data}
\label{sec:SK}

Super-Kamiokande (SK) is a water-Cherenkov detector with a fiducial volume of 22.5~ktons ($1.5 \times 10^{33}$ free protons and $7.5 \times 10^{32}$ oxygen nuclei). The best data in the energy range of neutrinos from PBH evaporation corresponds to that used in the search for the DSNB~\cite{Bays:2011si, Zhang:2013tua, Super-Kamiokande:2021jaq}, which is divided into four phases: SK-I ($t_{\rm I} = 1497$~days), SK-II ($t_{\rm II} = 794$~days), SK-III ($t_{\rm III} = 562$~days) and SK-IV ($t_{\rm IV} = 2970$~days). Indeed, limits on the integrated DSNB flux obtained with data from the first three phases~\cite{Bays:2011si} have been recently used to set bounds on the DM density fraction in the form of PBHs~\cite{Dasgupta:2019cae}. Nevertheless, the energy spectrum of the DSNB and that of neutrinos from PBH evaporation is not the same, so a detailed analysis must account for it. This is what we intend to do in this section.

The main detection channel of these neutrinos in SK is inverse beta-decay (IBD) ($\bar{\nu}_e \, p \rightarrow n \, e^+$), with subdominant contributions from $\nu_e$ and $\bar{\nu}_e$ charged-current (CC) interactions off oxygen nuclei and, in the absence of neutron tagging, from $\nu_\mu$ and $\bar{\nu}_\mu$ CC interactions producing muons below Cherenkov threshold. Without requiring the detection of a delayed neutron (by detecting the 2.2~MeV photon released after neutron capture in hydrogen), the lowest energies considered in these searches are mainly determined by the ability to remove the radioactive spallation products caused by cosmic-ray muons. The analyses of the first three SK phases allowed placing bounds on the integrated DSNB flux above $E_\nu \simeq E_e - (m_n - m_p) \simeq 17.3$~MeV (visible energy threshold at $E_e = 16$~MeV). However, by tagging neutrons in delayed coincidence with the positron signal, a significant suppression of the background can be achieved and the energy threshold can be lowered. With the new trigger system, which allows neutron tagging, this is the main goal of the SK-IV analysis. A first study using partial SK-IV data extended the search down to $E_\nu \simeq 13.3$~MeV ($E_e = 12$~MeV), with a total lifetime of 960~days~\cite{Zhang:2013tua}. Nevertheless, the additional selection cuts to reduce the probability of accidental background events resulted in an efficiency of finding delayed events of $17.74\%$, which correspondingly reduced the number of potential signal events. Given that most of the DSNB signal is expected to increase with decreasing energy, this analysis focused on the visible energy interval $E_e = (12 - 30)$~MeV, but yet, due to the limited statistics, the bounds obtained were weaker than with previous data, so these data had a minor effect on the overall results. Recently, the entire data set from SK-IV has been analyzed to extend the searches down to $E_\nu \simeq 9.3$~MeV ($E_e = 8$~MeV)~\cite{Super-Kamiokande:2021jaq}. Two new analyses have recently improved upon the one of the first three phases, by including two neutron tagging regions and a better spallation reduction algorithm. Yet, due to the preference for a small excess within this data set, the combined (SK-I, SK-II, SK-III, and SK-IV) limits are very similar to those obtained for the first three phases. In this work, we only include SK-I, SK-II, and SK-III data sets and we perform an analysis similar to the one of SK~\cite{Bays:2011si, Bays:2012wty}.

The two main sources of background at these energies are the Michel electrons and positrons from the decays at rest of invisible muons, which are produced below the threshold for Cherenkov emission by atmospheric $\nu_\mu$ or $\bar{\nu}_\mu$, and electrons and positrons produced by atmospheric $\nu_e$ and $\bar{\nu}_e$. Additionally,  neutral-current (NC) elastic events, which can produce de-excitation gammas, and low energy muons and pions, misidentified as electrons or positrons, also constitute other sources of background.\footnote{Note that the guaranteed, but uncertain, DSNB would also represent a background for the PBH signal. Below we comment on its potential impact on neutrino limits, but we do not include it in the current analysis.} It turns out that the average Cherenkov angle of photons from NC processes is large, whereas slow muons and pions emit Cherenkov light with a small angle. In contrast, IBD positrons are highly relativistic and emit Cherenkov light with an angle of $\sim 42^\circ$. The SK analysis of the first three data sets did not add a selection cut on the angle, but considered three Cherenkov angle regions: $20^\circ-30^\circ$ (`$\mu/\pi$' region), $38^\circ-50^\circ$ (`signal' region) and $78^\circ-90^\circ$ (`NC elastic' region).

As done in Ref.~\cite{Bernal:2012qh}, we include interactions of $\bar{\nu}_e$ off free protons, using the IBD differential cross section~\cite{Vogel:1999zy, Strumia:2003zx}, and we consider a relativistic Fermi gas model~\cite{Smith:1972xh} to include $\nu_e$ and $\bar{\nu}_e$ interactions off oxygen nuclei. In addition, we also include the PBH signal from $\nu_\mu$ and $\bar{\nu}_\mu$ CC interactions, with final muons below Cherenkov threshold (in analogy to the invisible muons background). The expected fraction of events in the visible electron/positron energy interval $E_{\rm vis} = [E_l,\, E_{l+1}]$, produced by neutrinos and antineutrinos from PBH evaporation, during phase $p$ = \{SK-I, SK-II, SK-III\} is given by
\begin{align}
\label{eq:evrateSK}
  ({\cal P}_{\rm PBH}^{p})_l & = A_{\rm PBH}^{p} \, \int  {\cal G}_{{p}, l} (E_e) \, dE_e  \nonumber\\
    & \times \Bigg\{ \int dE_\nu \, \left[ \left( \frac{d\sigma_{\rm f}^{\bar{\nu}_e}(E_\nu,E_e)}{dE_e} + \frac{1}{2} \, 
    \frac{d\sigma_{\rm b}^{\bar{\nu}_e}(E_\nu,E_e)}{dE_e}
    \right) \frac{d\Phi_{\bar{\nu}_e}}{dE_\nu} +
    \frac{1}{2} \, \frac{d\sigma_{\rm b}^{{\nu}_e}(E_\nu,E_e)}{dE_e}  \, \frac{d\Phi_{\nu_e}}{dE_\nu}  \right]  \nonumber \\
    & \quad + \int^{(E_\mu)_{\rm th}} dE_\mu \int dE_\nu \left[ \left( \frac{d\sigma_{\rm f}^{\bar{\nu}_\mu}(E_\nu,E_\mu)}{dE_\mu} + \frac{1}{2} \, \frac{d\sigma_{\rm b}^{\bar{\nu}_\mu}(E_\nu,E_\mu)}{dE_\mu} 
    \right) \frac{d\Phi_{\bar{\nu}_\mu}}{dE_\nu} \, \frac{dN_{\rm f}}{dE_e} \right. \nonumber \\
    & \left. \hspace{2.5cm} + \, \frac{1}{2} \, \frac{d\sigma_{\rm b}^{{\nu}_\mu}(E_\nu,E_\mu)}{dE_\mu} \, \frac{d\Phi_{\nu_\mu}}{dE_\nu} \left( \left(1 - \varepsilon_{\rm abs}\right) \, \frac{dN_{\rm f}}{dE_e} + \varepsilon_{\rm abs} \, \frac{dN_{\rm b}}{dE_e} \right) \right] \Bigg\},
\end{align} 
where $A_{\rm PBH}^{p}$ is a normalization constant so that $\sum_l ({\cal P}_{\rm PBH}^{p})_l = 1$. The cross sections for IBD and for CC interactions with nuclei are given by $\sigma_{\rm f}$ and $\sigma_{\rm b}$, respectively, and the factor $1/2$ accounts for water having twice as many free protons as oxygen nuclei. The Gaussian energy resolution of each SK phase, ${\cal R}_{p}(E_e,E_{\rm vis})$, is incorporated through ${\cal G}_{{p}, l} (E_e) = \int_{E_l}^{E_{l+1}} \, \epsilon_{p} (E_{\rm vis}) \, {\cal R}_{p}(E_e,E_{\rm vis}) \, dE_{\rm vis}$, with $\epsilon_{p}(E_{\rm vis})$ the detection efficiency, and where $E_e$ and $E_{\rm vis}$ are the true and detected electron/positron energy, respectively. The energy-dependent efficiency\footnote{Note that the efficiency is given as a function of the reconstructed energy, $E_{\rm vis}$, and not the true energy, $E_e$. The trigger efficiency is close to 100\% in the region of interest, so $\epsilon_{\rm SK}$ accounts for post-trigger reduction cuts to the data~\cite{Bays:2011si, Bays:2012wty}.}~\cite{Bays:2011si, Bays:2012wty} and the energy resolution~\cite{Hosaka:2005um, Cravens:2008aa, Abe:2010hy} of each SK phase were compiled in Ref.~\cite{Bernal:2012qh}. The second line represents the contribution from $\nu_e$ and $\bar{\nu}_e$ CC interactions, whereas the last two lines correspond to the $\nu_\mu$ and $\bar{\nu}_\mu$ events with muons below Cherenkov threshold ($E_\mu < (E_\mu)_{\rm th} \simeq 160$~MeV), produced by neutrinos with $E_\nu > 110$~MeV. For the latter contribution, we consider the spectrum of electrons/positrons from muon decays at rest ($dN_{\rm f}/dE_e$) and in orbit in an oxygen nucleus ($dN_{\rm b}/dE_e$)~\cite{Haenggi:1974hp}, accounting for a probability of absorption of $\mu^-$ of $\varepsilon_{\rm abs} = 0.184$.\footnote{In the analyses, for the contribution to the signal from $\nu_\mu$ and $\bar{\nu}_\mu$ CC interactions, we use the shape of the invisible muon background, for each phase, from Ref.~\cite{Bays:2012wty}, and as normalization we use the total number of invisible muon events obtained from the last two lines in Eq.~(\ref{eq:evrateSK}).} Due to the minimum required neutrino energy to produce these events, this contribution is more important for small PBH masses (for non-evaporated PBHs) and for broad mass distributions, whose flux extends to higher energies. For large masses, the PBH contribution from invisible muons is much smaller than that from $\nu_e$ and $\bar{\nu}_e$ CC interactions and can be neglected.

To describe the shape of the four backgrounds in each of the angular regions of observation, we use the probability distribution functions provided in Ref.~\cite{Bays:2012wty} and we fit their overall normalizations (with fixed relative normalizations among the three Cherenkov regions). The potential signal events from PBHs evaporation are assumed to only contribute to the `signal' region, which is a very good approximation. We perform a fit to the number of events of these five types: $N_{\rm PBH}^{p}$ (PBH signal), $N_{{\rm inv.} \mu}^{p}$ (invisible muons background), $N_{\nu_e {\rm CC}}^{p}$ (atmospheric $\nu_e$ background), $N_{\rm NC}^{p}$ (NC elastic background) and $N_{\mu/\pi}^{p}$ ($\mu/\pi$ background). We consider 18 4-MeV bins in the visible electron/positron energy interval $E_{\rm vis} = (16 - 88)$~MeV and perform an extended maximum likelihood analysis, with the likelihood function for each SK phase defined as 
\begin{equation}
\mathcal{L}_{p} = e^{-N_{\rm tot}^{p}} \,
\prod_{r=1}^{3} \, \prod_{l=1}^{18} \frac{\left[ \sum_i N_i^{p} \, ({\cal P}_i^{p})_l^r \right]^{\left(D_l^r\right)^{p}}}{\left(D_l^r\right)^{p}!} ~,
\end{equation}
where $r$ runs over the three Cherenkov regions, $l$ indicates the energy bin, $({\cal P}_i^{p})_l^r$ is the fraction of events of type $i = \{{\rm PBH}, \, {\rm inv.} \mu, \, \nu_e {\rm CC}, \, {\rm NC}, \, \mu/\pi \}$ in the $l^\text{th}$ bin and in region $r$, during phase $p$, $N_{\rm tot}^{p} = \sum_i N_i^{p}$ is the expected total number of events during phase $p$, and $\left(D_l^r\right)^{p}$ is the number of detected events in the $l^\text{th}$ bin in region $r$ during phase $p$.

\begin{figure}
	\centering 
	\includegraphics*[width=0.73\textwidth,trim=0 0 0 170]{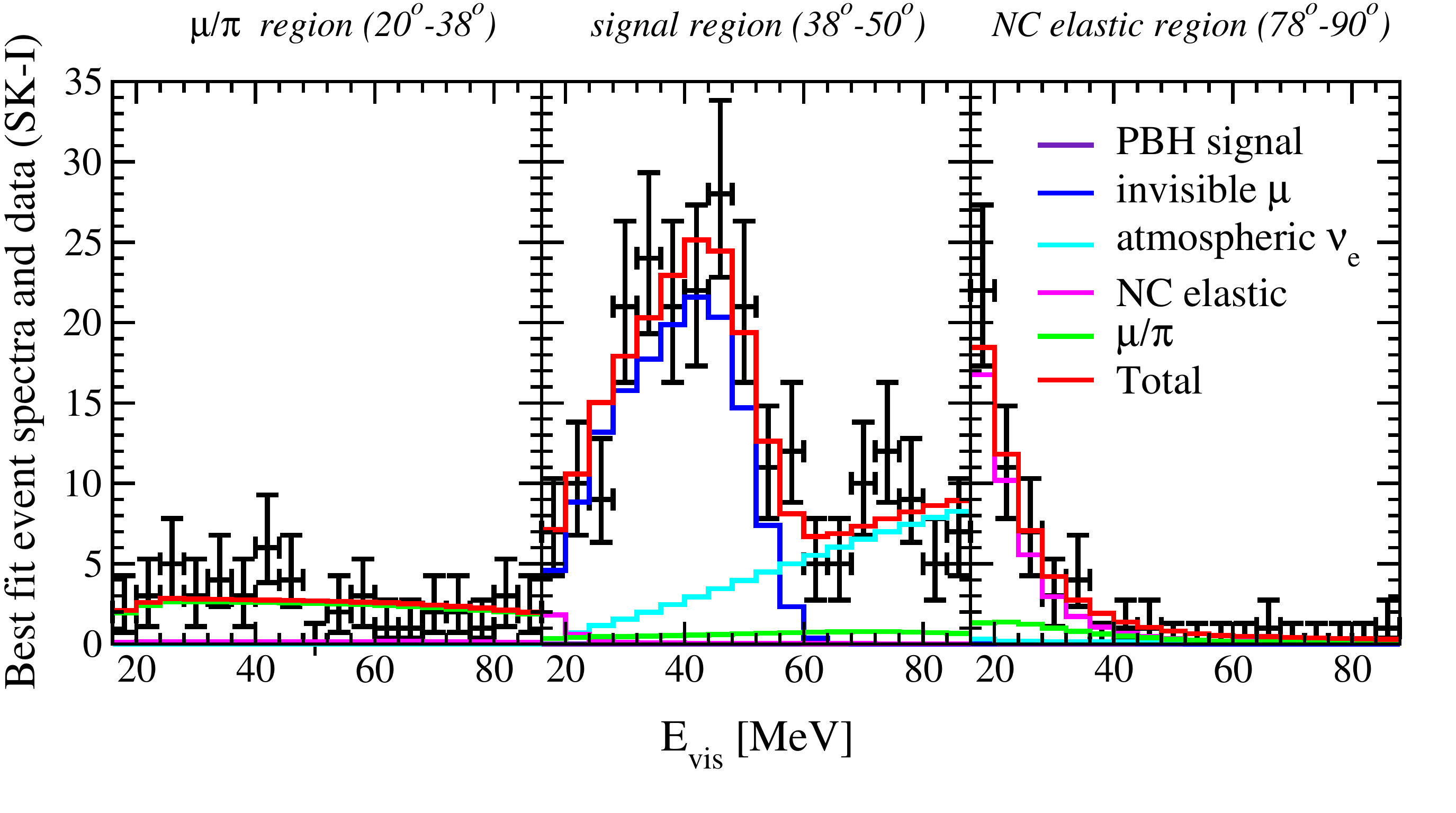} \\ 
	\includegraphics*[width=0.73\textwidth,trim=0 0 0 170]{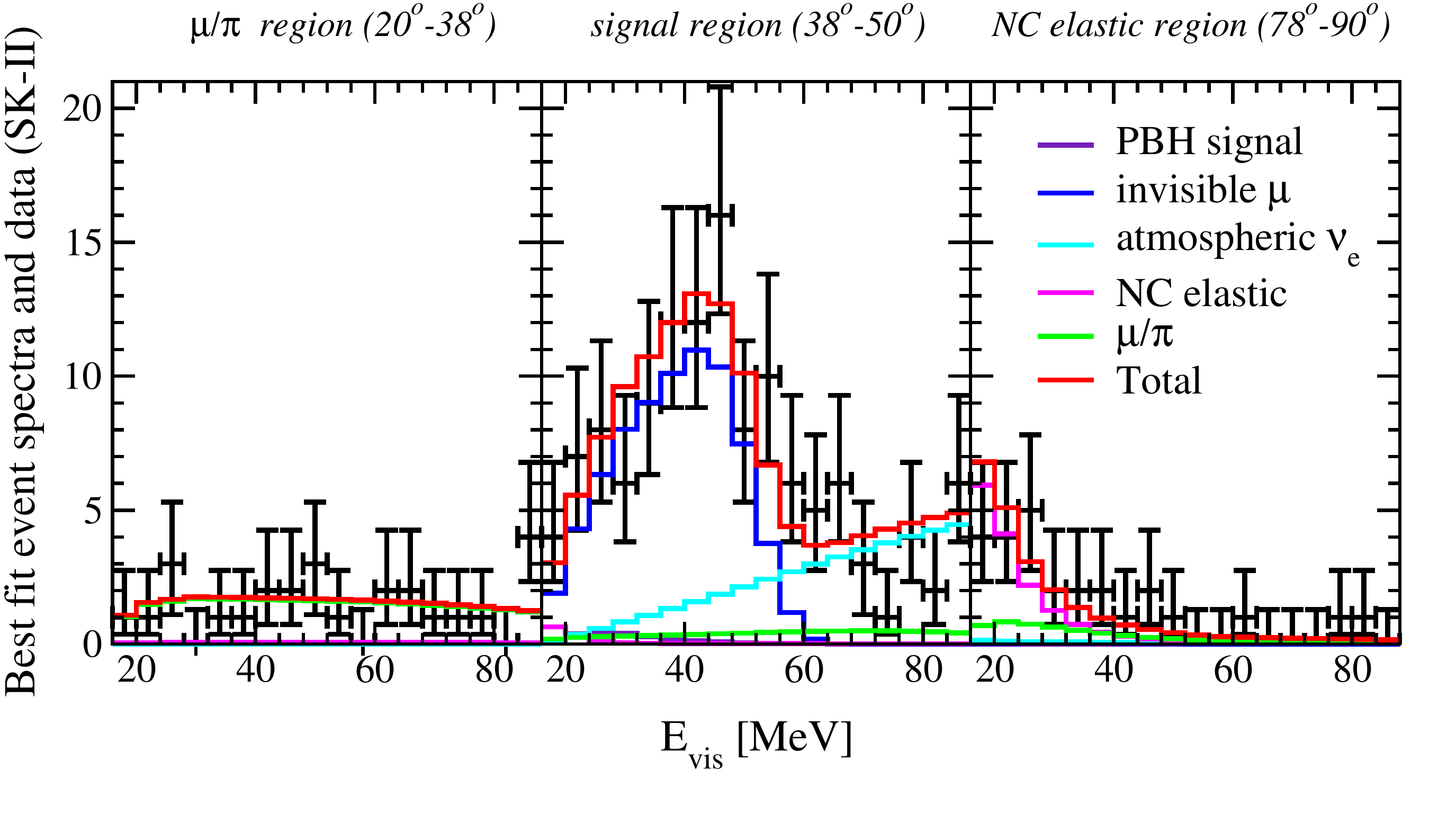} \\ 
	\includegraphics*[width=0.73\textwidth,trim=0 0 0 170]{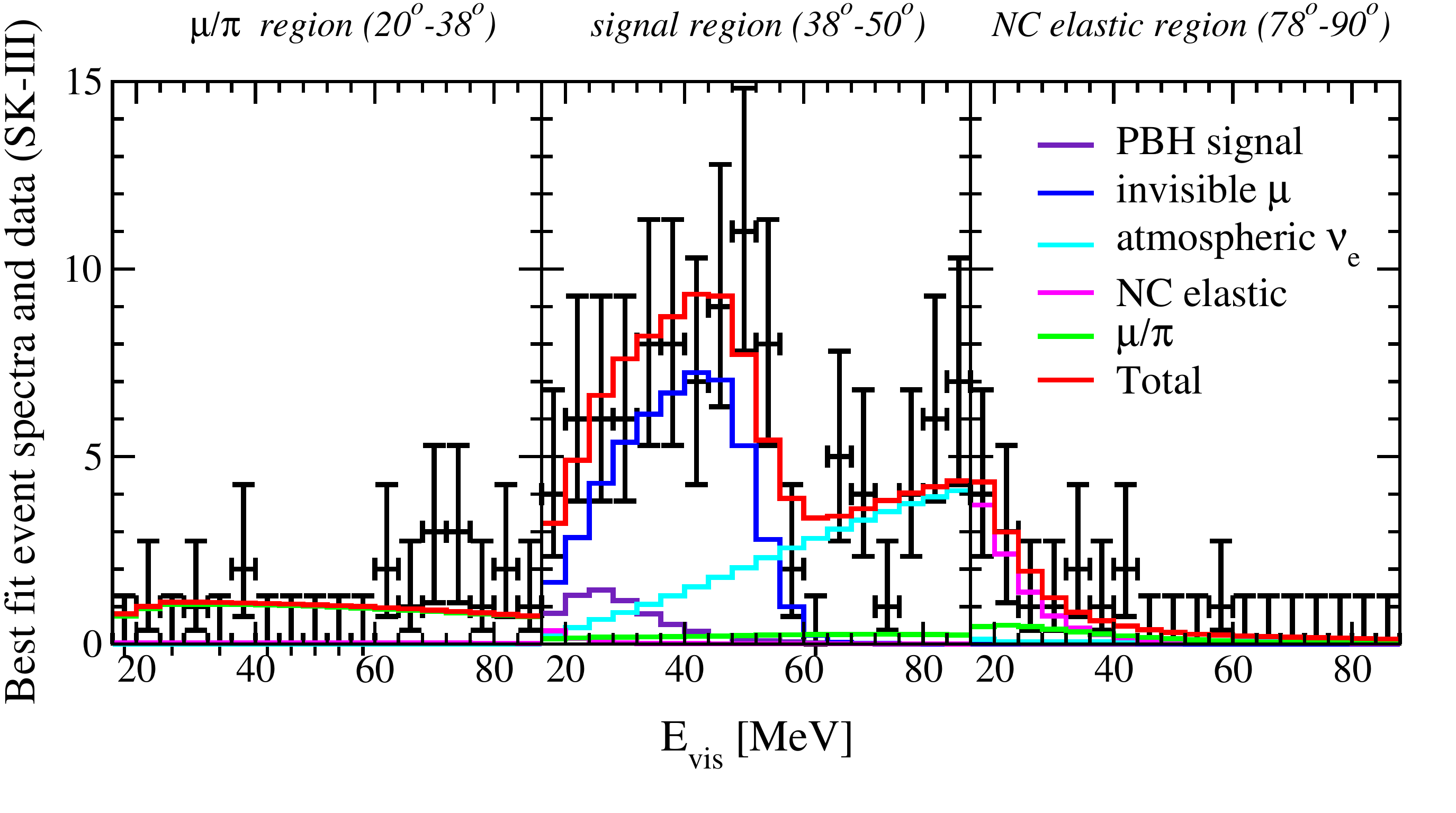}
	\caption{Best fit event spectra and data~\cite{Bays:2011si} for SK-I, SK-II and SK-III, in three Cherenkov angle regions, for $M_{\rm PBH} = 2 \times 10^{15}$~g and for a monochromatic mass distribution. We show the PBH signal spectra (violet histograms), the two main backgrounds in the signal region, invisible muons (blue histograms) and atmospheric $\nu_e$ and $\bar{\nu}_e$ (cyan histograms), and the $\mu/\pi$ (green histograms) and NC elastic (magenta histograms) backgrounds. We also show the total best fit spectra (red histograms). The best fits for the PBH signal correspond to $f_{\rm PBH} = 0$ (SK-I), $f_{\rm PBH} = 5.9 \times 10^{-4}$ (SK-II) and $f_{\rm PBH} = 2.3 \times 10^{-3}$ (SK-III), whereas the joint best fit is $f_{\rm PBH} = 0$.}
	\label{fig:SKevents}
\end{figure}

We do not include energy scale and energy resolution uncertainties, but we do account for the energy-independent systematic error on the efficiency. To do so, we slightly modify the likelihood above, following Refs.~\cite{Bays:2011si, Bays:2012wty}. All in all, the inclusion of these uncertainties only changes the results at the percent level or below. The total likelihood is maximized for each SK data set, treating the background normalizations as nuance parameters. We add a restriction on $N_{{\rm inv.} \mu}^{p}$ such that we allow a 30\% variation over the expected number invisible muon events using the FLUKA flux ($\sim 38$ events/year).\footnote{Note that the $\nu_\mu$ and $\bar{\nu}_\mu$ fluxes from Ref.~\cite{Honda:2015fha} (HKKM) are up to 10\% larger than the FLUKA flux for $E_\nu \lesssim 300$~MeV, so by allowing a 30\% (flat) variation over the FLUKA flux, we allow a $\sim 40\%$ downwards variation over the HKKM flux.} We also restrict $N_{\nu_e{\rm CC}}^{p}$ to guarantee its correlation with $N_{{\rm inv.} \mu}^{p}$, such that $N_{\nu_e{\rm CC}}^{p} = (0.4 - 0.8) \, N_{{\rm inv.} \mu}^{p}$ in the observation energy interval, in agreement with expectations. The likelihood for each phase is rescaled as a function of the number of signal events/year, ${\cal N}_{\rm PBH}$, and the total likelihood results from their product, $\cal{L}_{\rm tot} = \cal{L}_{\rm SK-I} \, \cal{L}_{\rm SK-II} \, \cal{L}_{\rm SK-III}$. The 90\% confidence level (CL) limit on the number of signal events/year, ${\cal N}_{\rm PBH, 90}$, is determined by solving
\begin{equation}
\frac{\int_0^{{\cal N}_{\rm PBH, 90}} \mathcal{L}_{\rm{tot}}
  ({\cal N}_{\rm PBH}) \, d{\cal N}_{\rm PBH}}{\int_0^{\infty}
  \mathcal{L}_{\rm{tot}} ({\cal N}_{\rm PBH}) \, d{\cal N}_{\rm PBH}} = 0.9\,.
\end{equation} 
The 90\% CL limit on the fraction $f_{\rm PBH}$ of PBHs contributing to the DM density is then given by 
\begin{equation}
f_{\rm PBH} (\Mbh) = \frac{{\cal N}_{\rm PBH, 90} (\Mbh)}{{\cal N}_{\rm PBH}^{\rm{tot}} (\Mbh)} ~, \qquad {\cal N}_{\rm PBH}^{\rm{tot}} \equiv \frac{\sum_{p} {\cal N}_{\rm PBH}^{p} \, t_{p}}{\sum_{p} t_{p}} ~, 
\end{equation}
where ${\cal N}_{\rm PBH}^{p}$ is the expected rate of signal events in the detection interval for each SK phase $p$, which corresponds to $f_{\rm PBH} = 1$. 

\begin{figure}
	\centering
	\includegraphics[scale=0.35]{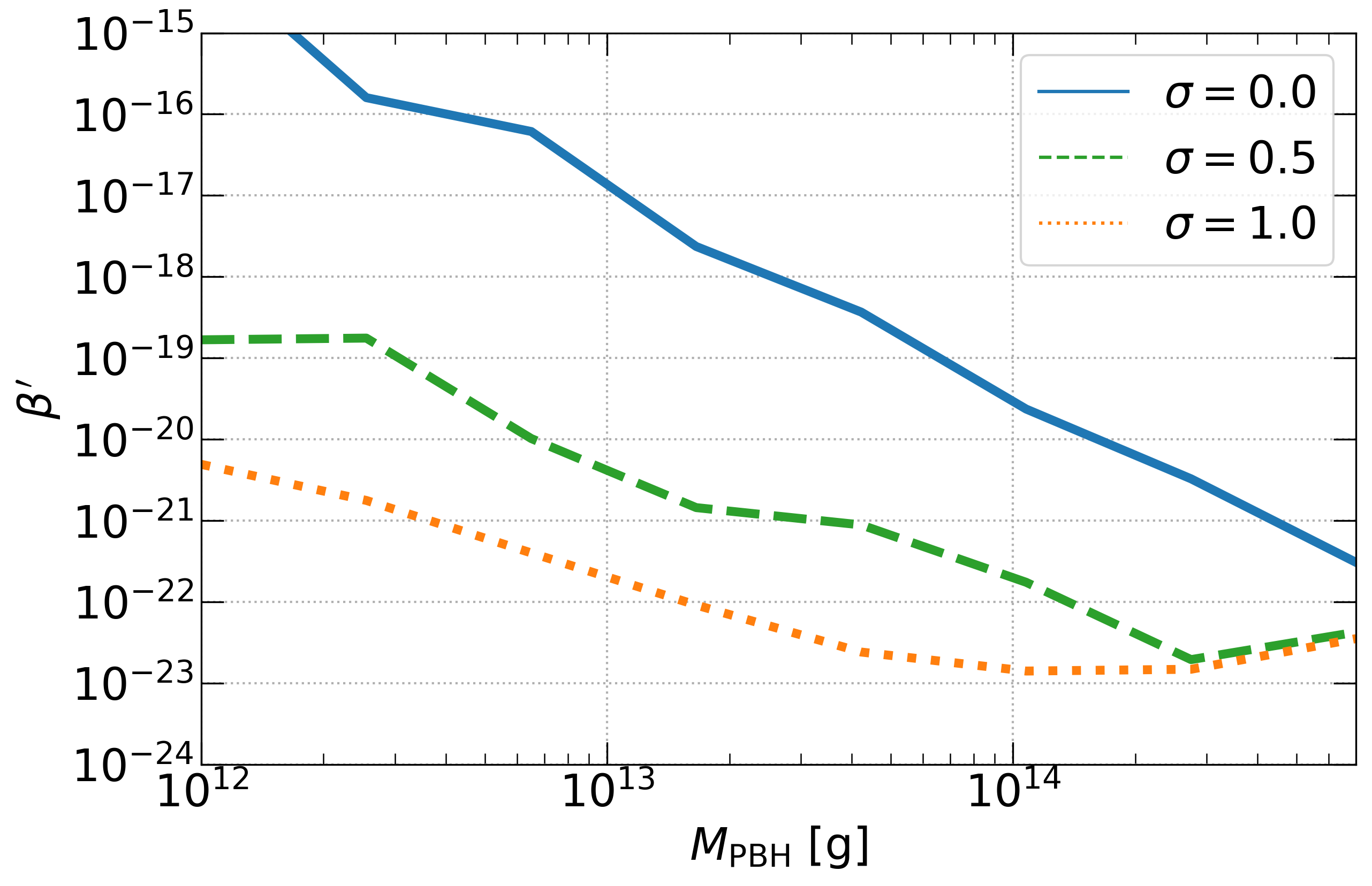}
	\includegraphics[scale=0.35]{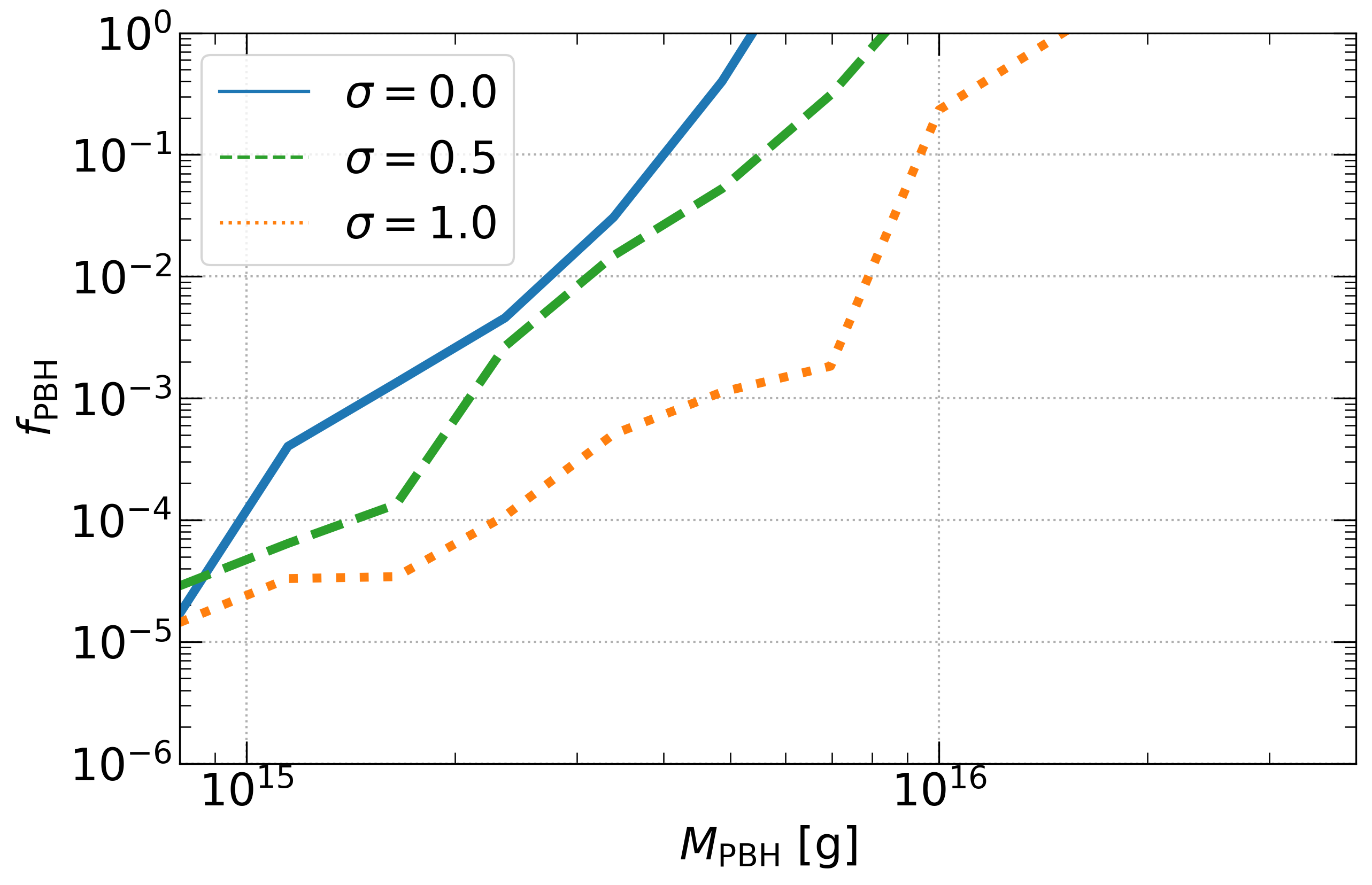}
	\caption{Updated bounds, at 90\% CL, from SK data on the PBH abundance as a function of the mass for monochromatic, $\sigma = 0$ (solid lines), and log-normal mass distributions, $\sigma = 0.5$ (dashed lines) and $\sigma = 1$ (dotted lines), in the case of fully evaporated (left panel) or non-evaporated (right panel) PBHs.}
	\label{fig:PBHbounds_SK}
\end{figure}

In Fig.~\ref{fig:SKevents}, we show the best fit for the case of $M_{\rm PBH} = 2 \times 10^{15}$~g with a monochromatic mass distribution. The best fits for the four backgrounds discussed above are depicted in the three Cherenkov regions. Note that for SK-II and SK-III, the best fit for the PBH signal is different from zero (barely visible for SK-II), $f_{\rm PBH} = 5.9 \times 10^{-4}$ (SK-II) and $f_{\rm PBH} = 2.3 \times 10^{-3}$ (SK-III). Yet, the joint best fit for the PBH signal for this PBH mass is $f_{\rm PBH} = 0$. This is not so for all PBH masses or the different mass distributions considered in this work, although we do not attempt to explain the observed signal in terms of PBHs evaporation, but just focus on setting limits on their abundance, as described above. The limits on the PBH abundance, as a function of the PBH mass, corresponding to the monochromatic ($\sigma = 0$) and log-normal ($\sigma = 0.5$ and 1) mass distributions, are shown in Fig.~\ref{fig:PBHbounds_SK}, for evaporated (left panel) and non-evaporated (right panel) PBHs. To ease the comparison with the projected bounds from future detectors, these SK limits will also be included in Figs.~\ref{fig:PBHbounds_DM}--\ref{fig:PBHbounds_lognormal_evap}.

Our results for the case of non-evaporated PBHs, that would account for part of the DM in the Universe, can be compared with the limits in Ref.~\cite{Dasgupta:2019cae} (similar to those in Ref.~\cite{Carr:2009jm}), which were obtained by imposing the integrated neutrino flux from PBHs evaporation to be lower than the allowed DSNB flux at SK. Here, as described above, we have performed a more complete analysis, using the spectral information of the detected events and accounting for the different energy distributions of signal and background events. This allows us to get more stringent limits than Refs.~\cite{Carr:2009jm, Dasgupta:2019cae}. We find that using the full spectral distribution becomes more important for smaller masses and for broader mass distributions. For a monochromatic mass distribution, this is particularly important for $\Mbh \lesssim 2 \times 10^{15}$~g. For these masses, the neutrino flux extends to much higher energies than in DSNB models, so the smaller the PBH mass the larger the ratio of the event rate to the integrated flux. This results in limits up to an order of magnitude more stringent than simply using the integrated DSNB flux. It turns out, however, that for $\Mbh \simeq 10^{15}$~g, current data allow a non-negligible contribution from neutrinos from PBHs evaporation, which weakens the would-be no-signal limit. This causes the small dent around those masses that can be seen in the right panel of Fig.~\ref{fig:PBHbounds_SK} (solid line). For $\Mbh \gtrsim 2 \times 10^{15}$~g, using the spectral information improves the constraints by less than a factor of two in the case of a monochromatic mass distribution. For extended mass distributions, accounting for the spectral information sets more stringent constraints up to larger masses (the larger the more extended the mass distribution). The improvement is of a factor of a few for the two log-normal mass distributions we consider and it is larger for the broadest one.
Nevertheless, even when making the comparison at the level of integrated fluxes, our results for extended mass distributions are different from those of Ref.~\cite{Dasgupta:2019cae}. For both cases, we obtain bounds which vary more pronouncedly with PBH mass than those in Ref.~\cite{Dasgupta:2019cae}, so that our limits are more constraining at small masses and less so at large masses. 
All in all, in comparison with Ref.~\cite{Dasgupta:2019cae}, our results differ by up to an order of magnitude for $\sigma = 0.5$ and up to two orders of magnitude for $\sigma = 1$, the largest differences occurring at the smallest PBH masses (with $\Mbh > M_{\rm evap}$).

The effect of considering signal events with muons produced below Cherenkov threshold also becomes more relevant for less massive non-evaporated PBH and for broader mass distributions, whose neutrino flux extends to higher energies where these events would mostly be produced. Counterintuitively, adding this contribution does not strengthen the constraints, but weakens them in general. This can be understood from the fact that the spectrum of invisible muons from neutrinos from PBHs evaporation is the same as that of invisible muons from atmospheric neutrinos, which represents the largest background. This degeneracy could allow accommodating a non-zero flux from PBHs, which would otherwise be disfavored. Yet, the constraint we have imposed on the contribution to invisible muons from the atmospheric neutrino flux does limit the contribution from PBHs. Finally, this addition only amounts to a correction of a few tens of percent in all cases. Had not we imposed such a constraint on the atmospheric neutrino contribution, the limits would (unrealistically) weaken by factors of up to four.

Additionally, we also compute limits on $\beta'$ for evaporated PBHs, extending to masses as low as $\Mbh = 10^{12}$~g. These limits represent an update and improvement over previous ones for a monochromatic mass distribution~\cite{Carr:2009jm}, which are about an order of magnitude less stringent than what we obtain. Furthermore, we also compute the limits for the two extended (log-normal) mass distributions considered above, which are more stringent the broader the mass distribution. As discussed above, the smaller PBH masses, the harder the emission spectrum, but also the earlier PBH evaporate and thus, the more substantial the energy redshift. Thus, the hardest neutrino flux at Earth corresponds to PBHs just evaporating now (see Fig.~\ref{fig:flux}), $\Mbh \simeq 8 \times 10^{14}$~g, and a qualitatively symmetric behavior of the limits for evaporating and non-evaporating PBHs is obtained.

\section{Event rates at future detectors}
\label{sec:rates}

\begin{figure}
	\centering
	\includegraphics[scale=0.25]{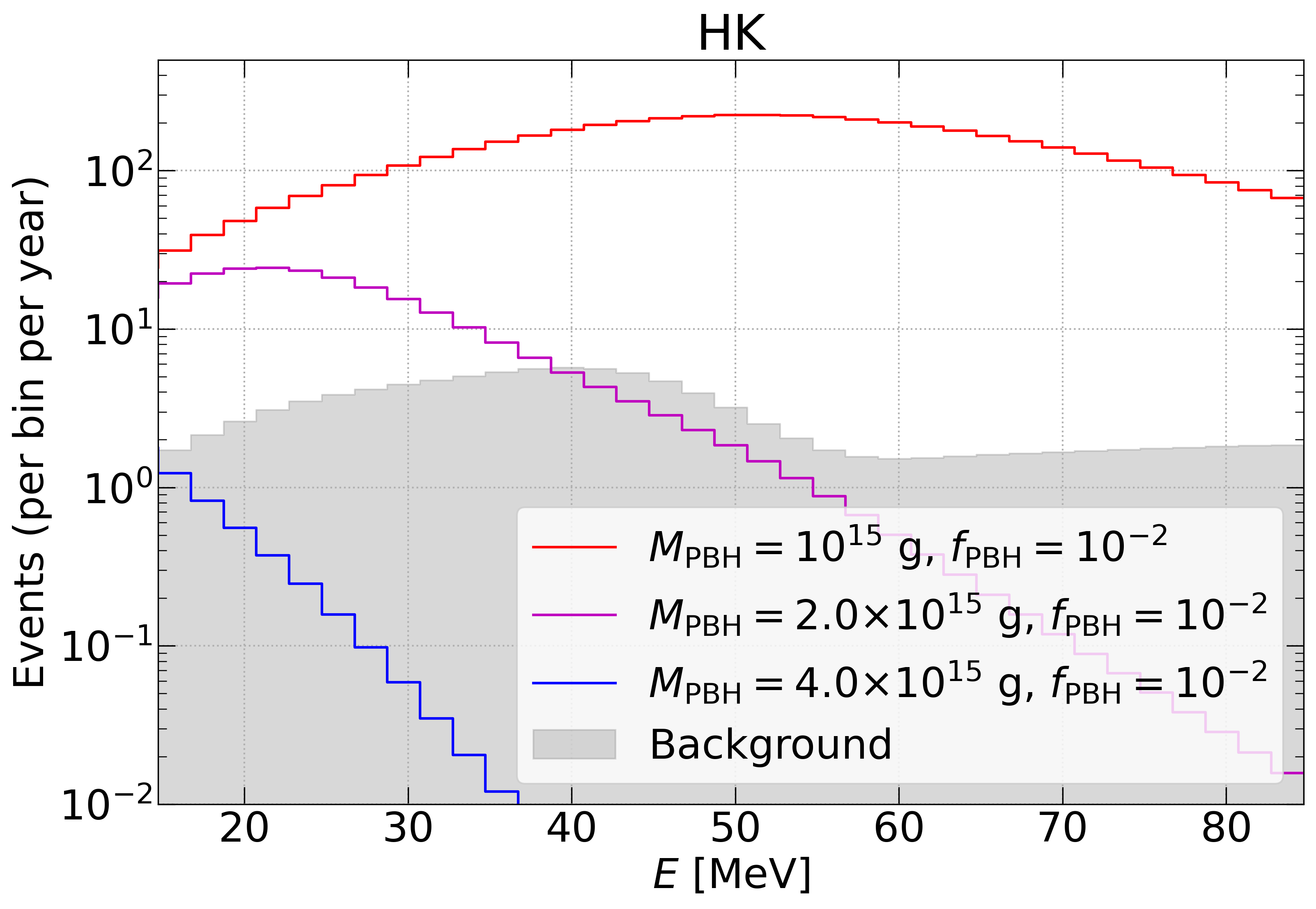}
	\includegraphics[scale=0.25]{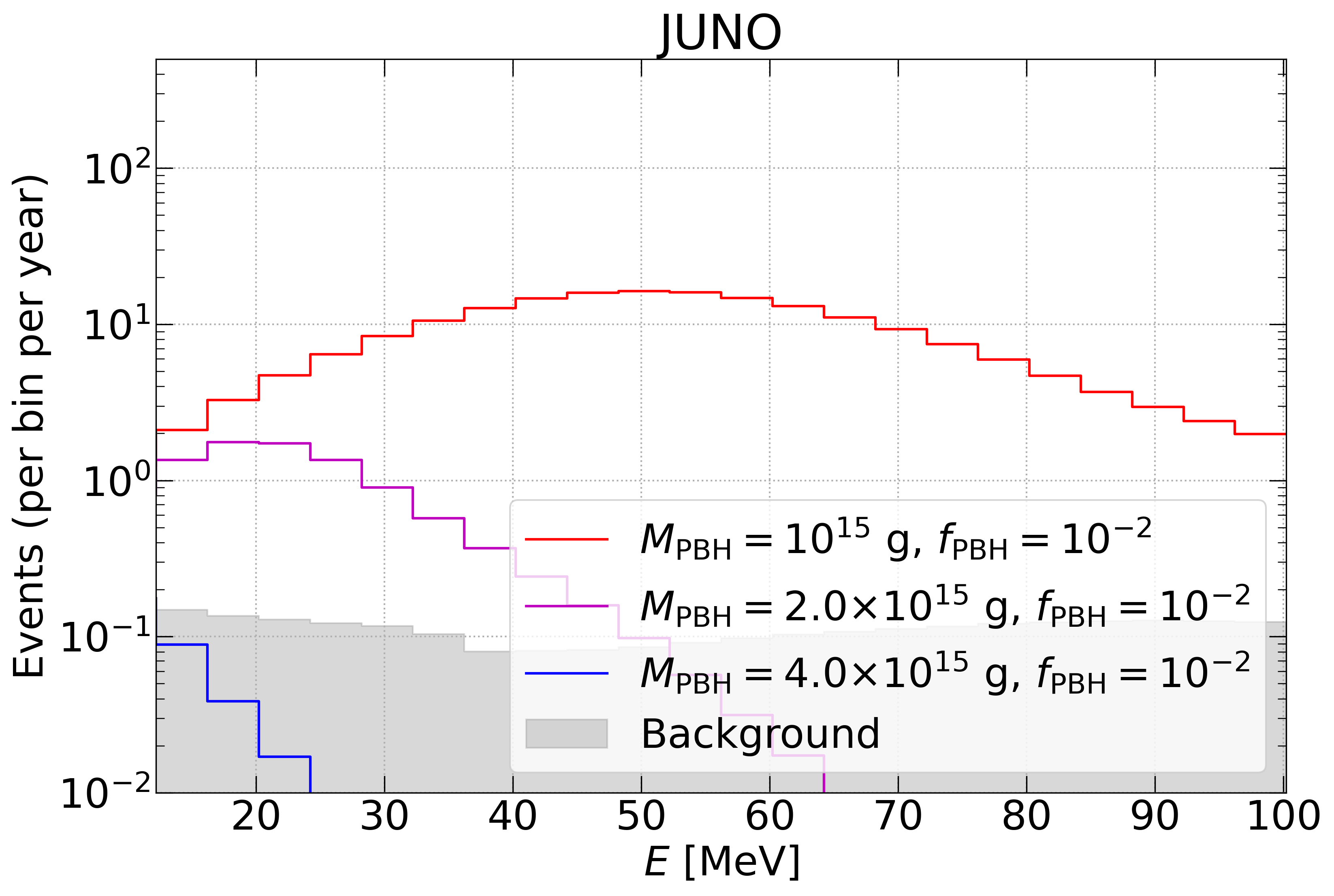}
	\includegraphics[scale=0.25]{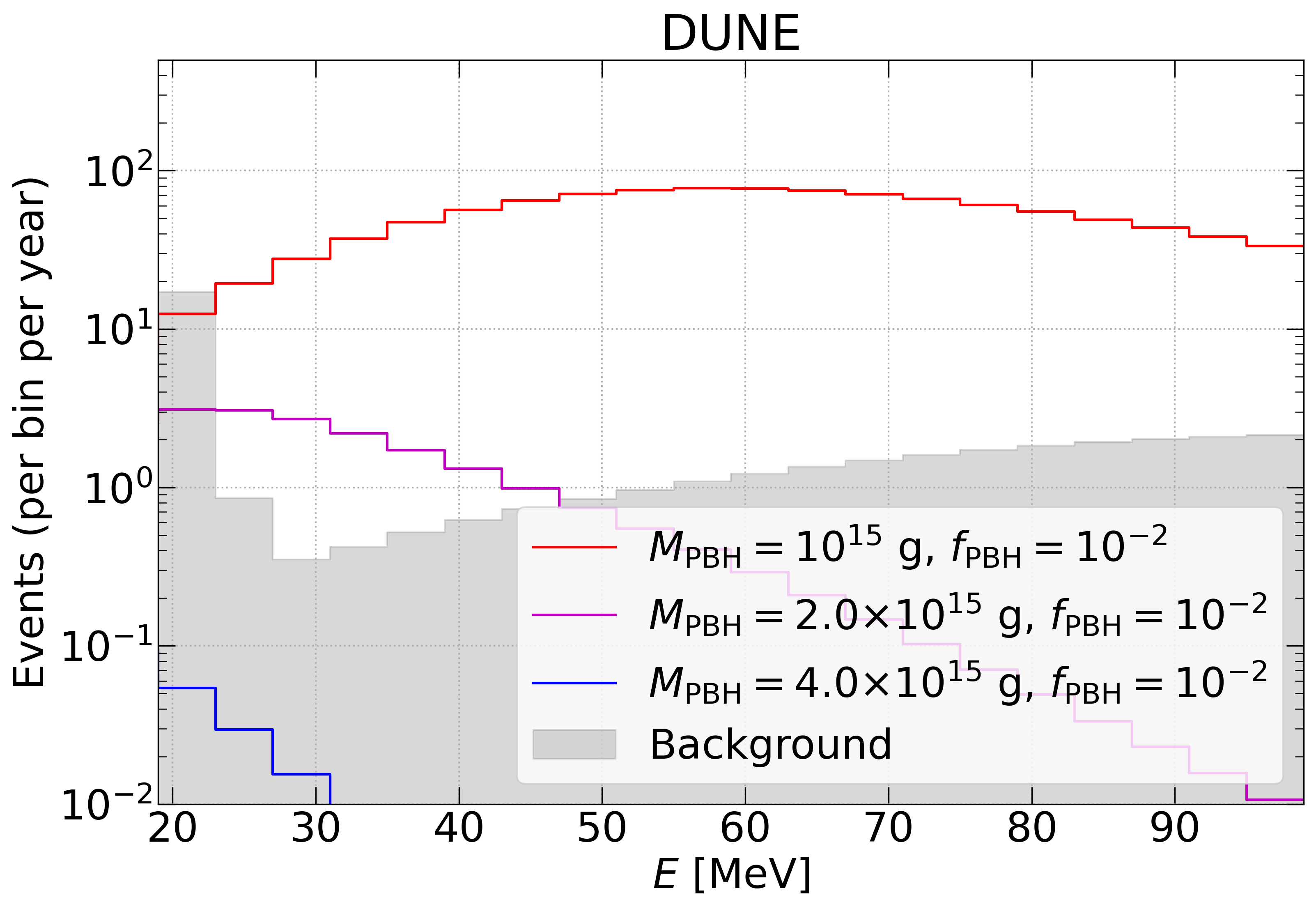}
	\includegraphics[scale=0.25]{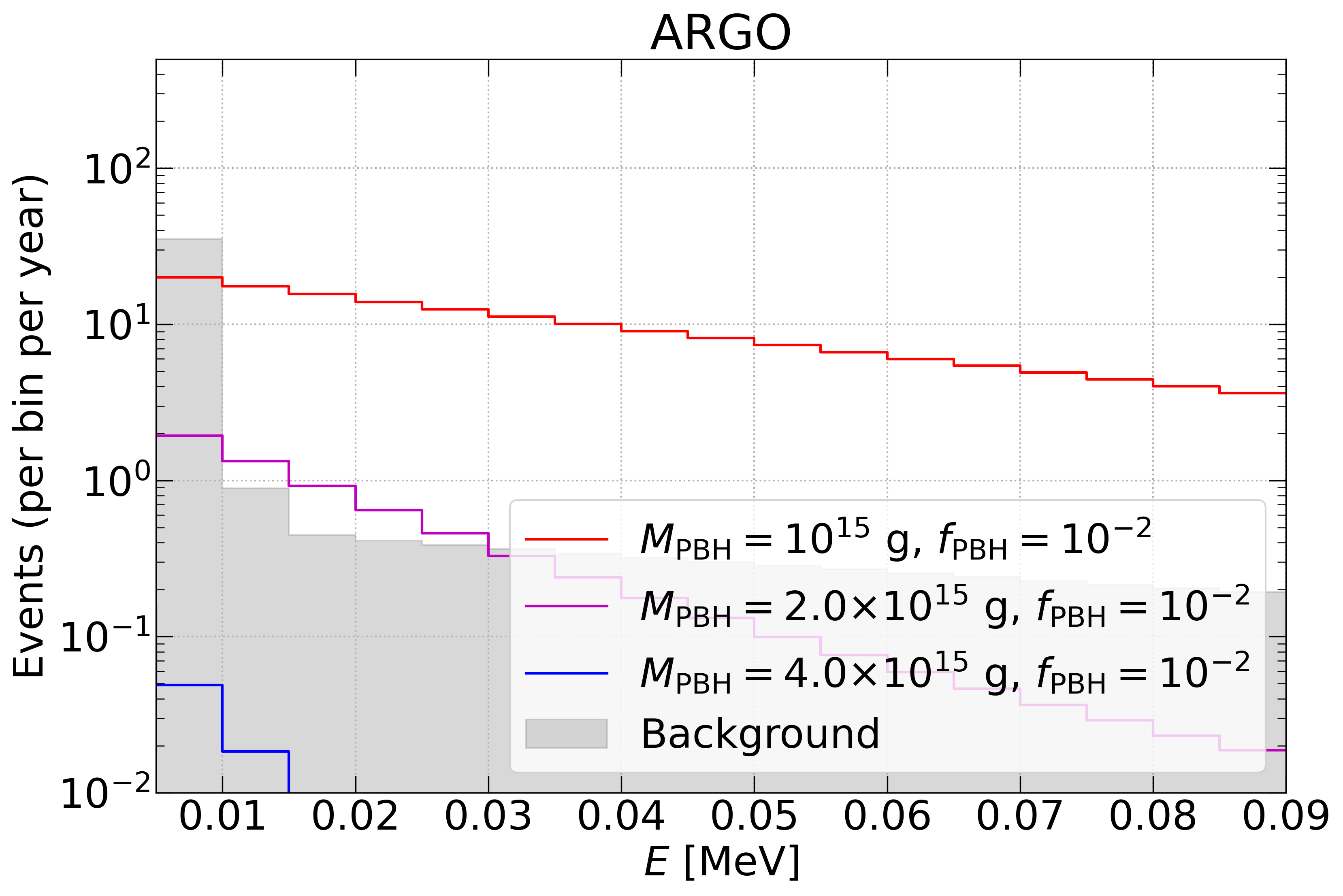}
	\caption{Event rates predicted for PBHs with three different masses, for a monochromatic mass distribution, as a function of the observed energy for several future detectors: Hyper-Kamiokande, JUNO, DUNE and ARGO. Binned rates are shown for $M_{\rm PBH} = 10^{15}$~g (red histograms), $M_{\rm PBH} = 2 \times 10^{15}$~g (purple histograms), and $M_{\rm PBH} = 4 \times 10^{15}$~g (brown histograms). In all cases, $f_{\rm PBH} = 10^{-2}$. The contribution from the main backgrounds (not including the DSNB) is indicated by the shaded regions. The low-energy tail of the background in DUNE and ARGO (lower panels) is caused by solar neutrinos, which have a negligible impact on HK and JUNO (upper panels).}
	\label{fig:events}
\end{figure}

In this section we consider future neutrino and DM detectors and describe the main detection channel of neutrinos (or antineutrinos) from PBHs in each case, as well as the most relevant backgrounds. A summary of some of the features of the detectors considered in this work is shown in Tab.~\ref{table:experiments}. The expected event rates for several PBH masses at different detectors are shown in Fig.~\ref{fig:events} along with the background rates (after experimental cuts). The code employed in this section to estimate the sensitivity reach for future experiments, {\tt nuHawkHunter}~\cite{pablo_villanueva_domingo_2022_6380867}, has been made publicly available on \href{https://github.com/vmmunoza/nuHawkHunter}{GitHub \faGithub}.\footnote{\url{https://github.com/vmmunoza/nuHawkHunter}}

\begin{table}
	\begin{center}
		\resizebox{\textwidth}{!}{
			\begin{tabular}{|c||c|c|c|c|}
				\hline
				Experiment & Type & Fid. mass (kt) & Main channel & Energy threshold (MeV) \\ 
				\hline\hline
				Super-Kamiokande \cite{Super-Kamiokande:2002weg}& Water Cherenkov & 22.5 & IBD & 16  \\
				\hline
				Hyper-Kamiokande~\cite{Abe:2018uyc} & Water Cherenkov & 374 & IBD & 16 \\
				\hline
				JUNO~\cite{Djurcic:2015vqa} & Liquid scintillator & 17 & IBD & 12 \\
				\hline
				DUNE~\cite{Abi:2020wmh} & Liquid argon & 40 & $\nu$Ar & 19  \\
				\hline
				DARWIN~\cite{Aalbers:2016jon} & Liquid xenon & 0.04 & CE$\nu$NS & 0.005 \\
				\hline
				ARGO\cite{Agnes:2020pbw} & Liquid argon & 0.37 & CE$\nu$NS & 0.005 \\
				\hline
			\end{tabular}
		}
	\end{center}
	\caption{Summary of the current and future detectors considered in this work, specifying their name, type, fiducial mass, main channel of detection and energy threshold for the analysis (visible energy for neutrino detectors and nuclear recoil energy for DM detectors). Only Super-Kamiokande is already operating, the others being at designing phases.}
	\label{table:experiments}
\end{table}

The different detectors we consider are most sensitive to different channels and present different peculiarities. Among the relevant neutrino interactions with nuclei, one has to consider CC interactions of the form $\nu_e + (N,Z) \rightarrow e^- + (N-1,Z+1)$ and $\bar{\nu}_e + (N,Z) \rightarrow e^+ + (N+1,Z-1)$, which lead to different products, such as de-excitation photons, leptons or neutrons. IBD belongs to the second category. On the other hand, NC include processes of the form $\nu + A \rightarrow \nu + A^*$, leading to the de-excitation of the nucleus, and also elastic scattering with nuclei, of the form $\nu + A \rightarrow \nu + A$. In general, independently of the specific process, for the forecasts presented in the following section for the different neutrino detectors, the event rates are approximately given by\footnote{An exception is the potential signal from muons produced below threshold in water-Cherenkov detectors. In this case, the spectrum is well defined and only the absolute number of events needs to be computed. In this section, though, we do not include this contribution. Note that, when doped with gadolinium, this background (and therefore, the analogous potential signal) is significantly reduced.}
\begin{equation}
    \frac{dR}{dE_{\rm vis}}\left(E_{\rm vis}\right) = \, N_t \int dE_\nu \, \epsilon \, \mathcal{R}\left(E_{\rm true}(E_\nu), E_{\rm vis}\right) \,  \frac{d\Phi_{\nu}}{dE_\nu}(E_\nu)\, \sigma(E_\nu) ~,
    \label{eq:eventrate}
\end{equation}
where $N_t$ is the number of targets, $\epsilon$ is the detection efficiency, $d\Phi_\nu/dE_\nu$ is the neutrino (or antineutrino) flux of the relevant flavor, $E_{\rm true}(E_\nu)$ is the true energy deposited in the detector, which is a function of the neutrino energy, $E_{\rm vis}$ is the visible energy, $\sigma$ is the detection cross section for the correspondent interaction, and $\mathcal{R}$ is the Gaussian resolution function,
\begin{equation}
\mathcal{R}\left(E_{\rm true}, E_{\rm vis}\right) = \, \frac{1}{\sqrt{2\pi} \,  \delta_{E}}\exp\left(-\frac{\left(E_{\rm true}-E_{\rm vis}\right)^2}{2 \, \delta_{E}^2}\right),
\label{eq:gauss}
\end{equation}
where $\delta_{E} (E_{\rm true})$ is the energy resolution. At sufficiently low energies, the deposited energy is approximately given in terms of the neutrino energy. For instance, in the case of IBD, we have checked that using the differential cross section instead, as done in the previous section, only introduces minor corrections at the few percent level in the event rate at the highest energies. Given that we provide forecasts of the sensitivity reach of future detectors and that we use a simplified description of them, we do not refine this calculation further.

\subsection{Hyper-Kamiokande}

Hyper-Kamiokande (HK)~\cite{Abe:2018uyc} is the expected successor of SK, designed to be built in Japan in the following decade. As its predecessor, it will be a water-Cherenkov detector, based therefore on the same physics, but with a larger volume. The planned final configuration contains two tanks of 187~kton of fiducial mass each, corresponding to $N_t = 2.5 \times 10^{34}$ targets in total. The results presented in this work assume 10 years of data with both tanks. 

As in other water-Cherenkov detectors, the main detection channel at HK is IBD, $\bar{\nu}_e + p \rightarrow n + e^+$. We use the cross section for the IBD channel tabulated in Ref.~\cite{Strumia:2003zx} (see also Ref.~\cite{Vogel:1999zy}).
At energies much below the proton mass, the positron energy, $E_e$, is approximately given in terms of the incoming neutrino energy, $E_\nu$, such that $E_{\rm true} = E_e \simeq E_\nu - (m_n - m_p) \simeq E_\nu - 1.3~\rm{MeV}$. 

The improved efficiency for neutron tagging by the addition of small amounts of gadolinium will help to significantly reduce the background and lower the energy threshold~\cite{Beacom:2003nk}.\footnote{After successfully developing the technology~\cite{Marti:2019dof}, the SK-Gd project was initiated in 2015, and in 2020, 10\% of the planned final Gd concentration was loaded~\cite{Super-Kamiokande:2021cuo}.} Although at the lowest energies, where spallation background cuts are critical to reduce the energy threshold, the efficiency is expected to be smaller than its value at energies $\gtrsim 20$~MeV, in this section the total efficiency is assumed to be energy-independent. Taking into account a 90\% efficiency of neutron capture (assuming a final concentration of 0.1\% of gadolinium) and a 74\% efficiency to detect 8~MeV photons, it is assumed to be $\epsilon=0.67$~\cite{Abe:2018uyc}. The energy resolution of the detector is taken into account through the integral over the energy resolution window $\mathcal{R}(E_e, E_{\rm vis})$, assumed to be a Gaussian profile with the same energy resolution of SK-III, $\left(\delta_E/E_e\right) = 0.0349 + 0.376/\sqrt{(E_e/{\rm MeV})} - 0.123 \, ({\rm MeV}/E_e)$~\cite{Abe:2010hy}.

Assuming neutron tagging, the main backgrounds for energies $\lesssim 20$~MeV are accidental coincidences with spallation products, spallations products with accompanying neutrons and the tail of reactor antineutrinos (important below $\sim 12$~MeV)~\cite{Abe:2018uyc}. Among the spallation products, the only one with an accompanying neutron is $^9$Li, but its short lifetime would allow reaching a very high rejection efficiency. After cuts, above 16~MeV these backgrounds are expected to be negligible, so this is the energy threshold we consider. As discussed above, at higher energies, Michel electrons/positrons from invisible muons and $\nu_e$ and atmospheric $\bar{\nu}_e$ CC events are the main backgrounds. However, the identification of IBD events by tagging neutrons in coincidence with the positron signal, would allow to significantly reduce the atmospheric backgrounds. In this work, we assume a factor of five background reduction with respect to the case of no gadolinium doping. For this background, we adopt the spectral shape corresponding to SK-I from Appendix~B of Ref.~\cite{Bays:2012wty}, normalizing it according to Ref.~\cite{Abe:2018uyc} (right panel of Fig.~188). For the atmospheric $\nu_e$ and $\bar{\nu}_e$ background, we make use of the fluxes from the Monte Carlo FLUKA simulations~\cite{Battistoni:2005pd}. The expected event rates induced by neutrinos from PBH evaporation (for several PBH masses and assuming a monochromatic mass distribution) and the background rate are shown in the upper left panel of Fig.~\ref{fig:events}.

\subsection{JUNO}

The Jiangmen Underground Neutrino Observatory (JUNO)~\cite{Djurcic:2015vqa, An:2015jdp} is a planned 20~kton liquid scintillator detector to be built in Jiangmen, China, consisting of a central tank filled with of linear alkylbenzene. The central detector is surrounded by a water-Cherenkov detector within a muon tracker, aimed at reducing certain sources of background, which will result in 17~kton of fiducial mass.

As in the case of water-Cherenkov detectors, the most important detection channel at the relevant energies is IBD~\cite{Lu:2016ipr, Wang:2020uvi}. Therefore, the event rate is also approximately given by Eq.~\eqref{eq:eventrate}, with $N_t = 1.2 \times 10^{33}$ targets. We assume a detection efficiency of $\epsilon = 0.5$, due to the application of pulse-shape discrimination techniques to suppress backgrounds~\cite{Mollenberg:2014pwa} (see below). Unlike Cherenkov detectors, rather than detecting the positron from the IBD reaction, liquid scintillator detectors tag photons from the subsequent positron-electron annihilation. Thus, the observed energy is given by $E_{\rm true} = E_e + m_e \simeq E_\nu - (m_n - m_p) + m_e \simeq E_\nu - 0.782$~MeV. Hence, $dR/dE_{\rm true} = dR/dE_e|_{E_{\rm true} = E_e + m_e}$. The width of the Gaussian energy resolution function for JUNO is assumed to be $\left(\delta_E/E_{\rm true}\right) \simeq 0.03/\sqrt{(E_{\rm true}/{\rm MeV})}$~\cite{An:2015jdp}.

The most relevant backgrounds include reactor antineutrinos, NC atmospheric neutrino and antineutrino events, and CC atmospheric antineutrino events. The reactor antineutrino background dominates at energies $E_\nu \lesssim 12$~MeV, so in our analysis we only consider $E_{\rm vis} \geq 12$~MeV. Above this energy and up to $\sim 40$~MeV, the dominant background (after selection cuts) is caused by NC interactions with $^{12}$C, which can produce free neutrons and other nuclei as final states, which later decay and mimic the IBD signature. We consider the estimate of the event rate corresponding to this background from Ref.~\cite{An:2015jdp} (right panel of Fig.~5-2) after the application of pulse-shape discrimination~\cite{Mollenberg:2014pwa}, which is based on the computations for the KamLAND~\cite{Collaboration:2011jza} and LENA~\cite{Mollenberg:2014pwa} detectors, that result in a signal efficiency of $\epsilon = 0.5$. At higher energies, the NC background can be neglected with respect to interactions of atmospheric $\bar{\nu}_e$ via CC (mainly IBD), which are the major source of background. To compute this background we consider the atmospheric neutrino fluxes obtained by the Monte Carlo FLUKA simulations~\cite{Battistoni:2005pd}, also assuming an efficiency of $\epsilon =0.5$. Since the tabulated atmospheric fluxes are computed at the Kamioka latitude, we correct them to account for latitude effects by a constant multiplying factor, interpolating between the latitude correction factors from Ref.~\cite{Wurm:2007cy}. Unlike in water-Cherenkov detectors, in liquid scintillator detectors, muons can be very efficiently tagged, so the invisible muon background is negligible. Finally, fast neutrons from cosmic muon decays represent another source of background. However, such events are expected to occur at the edges of the detector, and thus its rate can be reduced to $\sim 1$~events/yr by introducing a soft fiducial volume cut. The application of pulse shape discrimination can further reduce this background down to $\sim 0.01$~events/yr~\cite{An:2015jdp}, making it negligible.

The expected event rates induced by neutrinos from PBH evaporation (for several PBH masses and assuming a monochromatic mass distribution) and the background rate are shown in the upper right panel of Fig.~\ref{fig:events}.

\subsection{DUNE}

The Deep Underground Neutrino Experiment (DUNE)~\cite{Abi:2020wmh, Abi:2020evt} includes a liquid argon detector to be built in South Dakota, USA, composed of four time projection chambers of 10~kton each, resulting in a fiducial volume of 40~kton. The most relevant detection channel is the CC interaction of electron neutrinos with argon,
\begin{equation}
\label{eq:DUNEch}
    \nu_e + ^{40}{\rm Ar} \rightarrow e^- + ^{40}{\rm K}^* ~.
\end{equation}

The event rate of a particular signal is computed via Eq.~(\ref{eq:eventrate}), with $N_t=6.02 \times 10^{32}$, and an efficiency $\epsilon=0.86$ (given by a trigger efficiency of $\sim$90\% and a reconstruction efficiency of 96\%~\cite{Ankowski:2016lab, Moller:2018kpn}). The total interaction cross section, $\sigma_{\rm Ar}(E)$, is obtained from Ref.~\cite{GilBotella:2003sz} (see also Refs.~\cite{Cocco:2004ac, Capozzi:2018dat}), and can be approximately fitted by
\begin{equation}
    {\rm log}_{10} \left( \frac{\sigma_{\rm Ar}(E_\nu)}{10^{-44} \, {\rm cm }^2} \right) \simeq -0.8  \, \left( \frac{E_\nu}{\rm MeV} \right)^{-0.035},
\end{equation}
which matches the tabulated values within $\sim 10 \%$ in the energy range $E_\nu = (3 - 100)$~MeV. Given that its uncertainty is energy dependent and not smaller than 10\%~\cite{Capozzi:2018dat}, this seems a reasonable fit. In our calculations, though, we use the tabulated values. Finally, the neutrino energy resolution ($E_{\rm true} = E_\nu$) for the DUNE detector is taken to be $\left(\delta_E/E_\nu\right) = 0.2$~\cite{Abi:2020evt} (right panel of Fig.~7.5).\footnote{Note that a more optimistic neutrino energy resolution for the DUNE detector has also been considered within the context of neutrinos from PBHs evaporation~\cite{DeRomeri:2021xgy}. Nevertheless, that energy resolution ($\sim 5\%$) actually corresponds to electrons and does not account for energy losses to neutrons, which limit the neutrino energy resolution to be above 15\% for $E_\nu > 20$~MeV~\cite{Abi:2020evt}.} 

At energies of a few MeV, the major sources of background in DUNE include neutron capture processes, due to radiogenic and cosmogenic sources, beta decays from atmospheric muon-induced spallation products, and $^8$B and \textit{hep} solar neutrinos (see, e.g., Ref.~\cite{Capozzi:2018dat}). Above $\sim 15$~MeV, solar and spallation backgrounds are the most important ones. Nevertheless, after applying some experimental cuts, the spallation background can be reduced to a negligible level at those energies~\cite{Zhu:2018rwc}. Due to the high solar neutrino flux, the neutrino energy resolution of the detector determines the actual neutrino energy threshold for new signals. This can be clearly seen in the lower left panel of Fig.~\ref{fig:events}, where we depict the expected PBH signal (for several PBH masses and assuming a monochromatic mass distribution) and background event rates. The minimum energy for signal events to overcome backgrounds is larger than the endpoint of the \textit{hep} solar neutrino spectrum (18.8~MeV). This impacts the maximum PBH mass that could actually be tested with DUNE. At higher energies, the dominant background comes from atmospheric $\nu_e$ CC interactions~\cite{Cocco:2004ac}, Eq.~\eqref{eq:DUNEch}.
To compute these backgrounds, we consider solar neutrino fluxes~\cite{Robertson:2012ib} and we use the FLUKA simulations for low-energy atmospheric neutrino fluxes~\cite{Battistoni:2005pd}, corrected by a multiplicative latitude factor~\cite{Wurm:2007cy}, as done above.

\subsection{Dark matter detectors: DARWIN and ARGO}

Direct DM detection experiments can also be effective neutrino detectors, via CE$\nu$NS. This process can occur as long as the wavelength of the momentum transferred to the target is large enough so that the entire region within the wavelength contributes to the amplitude.

Among the several projected DM detectors, DARWIN~\cite{Aalbers:2016jon} will be composed of liquid xenon, and ARGO~\cite{Agnes:2020pbw} of liquid argon, both planned to be placed at Gran Sasso, Italy. Here, for both experiments we assume a threshold for the nuclear recoil energy of $(E_{\rm nr})_{\rm th} = 5$~keV and a maximum of $(E_{\rm nr})_{\rm max} = 100$~keV, but we do not explicitly describe the quenching effects which could alter the energy spectrum. Note that $(E_{\rm nr})_{\rm th}$ is not the physical threshold of the experiments. Below this energy, however, the solar neutrino background would bury any possible signal from PBHs evaporation. 

In the absence of physics beyond the SM, the CE$\nu$NS cross section with a nucleus $A$, neglecting corrections of order ${\cal O}(E_\nu/m_A)$ and ${\cal O}(E_{\rm nr}/E_\nu)$, is given by~\cite{Freedman:1977xn}
\begin{equation}
\dfrac{d \sigma_A (E_{\nu}, E_{\rm nr})}{d E_{\rm nr}} \simeq {\cal Q}_W^2 \, \dfrac{G_F^2 \, m_A}{2 \, \pi} \left(2 - \dfrac{m_A \, E_{\rm nr}}{E_{\nu}^2}\right) F_{A}^2 (E_{\rm nr}) ~,
\label{eq:nucxsec}
\end{equation}
where $E_{\rm nr}$ is the recoil energy of the target nucleus, $m_A$ is the mass of the target nucleus, $G_F$ is Fermi coupling constant, ${\cal Q}_W = \left( N_A \, g_{V, n} + Z_A \, g_{V,p} \right)$ is the weak hypercharge of a nucleus with $N_A$ neutrons and $Z_A$ protons, and the weak charges of the neutron and the proton are $g_{V, n} = -1/2$ and $g_{V,p} = 1/2 - 2 \sin^2 \theta_W$, with $\theta_W$ is the weak mixing angle, and $F_A$ is the nuclear form factor, taken to be the Helm form factor~\cite{Helm:1956zz, Lewin:1995rx}.\footnote{Note that $F_A = \frac{1}{{\cal Q}_W} \left[N \, F_n - Z \, \left(1 - 4 \sin^2 \theta_{\rm w}\right) \, F_p \right]$, where $F_n$ and $F_p$ are the neutron and proton form factors, which in this work, are assumed to be equal.} Since at low momentum transfer $\sin^2 \theta_W = 0.23857$~\cite{Zyla:2020zbs}, the CE$\nu$NS cross section scales approximately as $N_A^2$, which results in a significant enhancement for heavy nuclei.

Unlike the sensitivity studies above, here we consider the differential scattering cross section, so the event rate produced by neutrinos scattering off a target nucleus $A$ is written as
\begin{equation} \label{eq:DM_rec_sig}
\dfrac{d R}{d E_{\rm vis}}(E_{\rm vis}) =  N_t \, \int d E_{\nu} \, d E_{\rm nr} \, \epsilon(E_{\rm nr}) \, \mathcal{R}(E_{\rm nr},E_{\rm vis}) \, \frac{d\Phi_{\nu}}{d E_\nu} (E_{\nu}) \, \dfrac{d \sigma_A (E_{\nu}, E_{\rm nr})}{d E_{\rm nr}} \, \theta(E_{\rm nr}^{\rm max} - E_{\rm nr}),
\end{equation}
where $E_{\rm nr}^{\rm max} = 2 \, E_\nu^2/(m_A + 2 \, E_\nu)$. We assume perfect efficiency, $\epsilon(E_{\rm nr}) = 1$, and an energy resolution ($E_{\rm true} = E_{\rm nr}$) given by $\left(\delta_E/E_{\rm nr}\right) = 0.32/\sqrt{(E_{\rm nr}/{\rm keV})} + 0.06$. This resolution approximately corresponds to expectations for an S2-based signal~\cite{Schumann:2015cpa}, being about 20\% at 5~keV in a xenon detector~\cite{Aalbers:2016jon}. Combining light and charge signals could improve the energy resolution by some tens of per cent, so from this point of view, our assumption is conservative. In this work, we use the same energy resolution for DARWIN and ARGO, although notice that liquid argon detectors could have a better energy resolution~\cite{Agnes:2020pbw}. The number of targets is $N_t = 1.8 \times 10^{29}$ (40~t) for DARWIN and $N_t = 5.7 \times 10^{30}$ (370~t) for ARGO.

For the analysis, we use a bin size of $\Delta E_{\rm vis} = 5$~keV in the measured (corrected for quenching) nuclear recoil energy. Notice that since the detection channel is a neutral-current process, all neutrino (and antineutrino) flavors contribute to the expected event spectrum. For these DM detectors, we consider the dominant backgrounds to be the solar and atmospheric neutrino fluxes (using the same fluxes as above) interacting with the target material via the CE$\nu$NS channel. The expected event rates induced by neutrinos from PBH evaporation (for several PBH masses and assuming a monochromatic mass distribution) and the background rate for ARGO are shown in the lower right panel of Fig.~\ref{fig:events}.

\section{Sensitivity reach of future detectors}
\label{sec:sensitivity}

Next, we present sensitivity studies by performing a binned extended likelihood analysis for each detector. In all cases, we assume a data collection time of $\mathcal{T} = 10$~years and a bin size of $\Delta E_{\rm vis} = 2$~MeV for HK, $\Delta E_{\rm vis} = 4$~MeV for JUNO and DUNE and $\Delta E_{\rm vis} = 5$~keV for the DM detectors (DARWIN and ARGO). We consider the maximum likelihood ratio, which for independently Poisson distributed data, can be written as 
\begin{equation}
    \chi^2 = 2 \, \sum_{i} \left[S_i + B_i - D_i+D_i \ln\left(\frac{D_i}{S_i + B_i}\right)\right],
\label{eq:chi2}
\end{equation}
where $S_i$ is the number of expected events in bin $i$ produced by neutrinos from PBHs evaporation, $B_i$ is the number of expected background events in bin $i$, and $D_i$ is the number of data points in bin $i$. For the forecasts, our null hypothesis is the case with no contribution from PBH evaporation, and thus we take $D_i = B_i$. Moreover, throughout the work, we assume the likelihood ratio to follow a $\chi^2$ distribution with one degree of freedom.

Note that the procedure we follow relies on several simplifying assumptions, which are expected to modify the sensitivities to different extent, although not dramatically. We have not included systematic uncertainties, as for instance in the background expectations, where we have not included the DSNB. Detection efficiencies have been taken as constants, while more realistic treatments would imply energy dependent efficiencies. Similarly, the energy resolution functions have been taken from the experimental proposals and may vary for the actual experiments. Furthermore, we have restricted ourselves to the main detection channels for each future experiment. Nonetheless, in detectors like HK and JUNO, other detection channels besides IBD could give rise to non-negligible (although subdominant) signals from PBHs, as well as their corresponding backgrounds. We have only considered Majorana neutrinos, while slightly different constraints would be obtained for Dirac neutrinos~\cite{Lunardini:2019zob}. Although detected events are not sensitive to the Dirac/Majorana nature, the existence of Dirac neutrinos would lead to PBHs emitting more degrees of freedom (12 rather than 6), and hence evaporating somewhat earlier and with a slightly different neutrino spectrum. It is also worth to note that all these results apply to PBHs in asymptotically flat space-time. Properly describing PBHs with a metric embedded in a cosmological space-time may notably change the bounds from Hawking evaporation~\cite{Picker:2021jxl, Xavier:2021chn}. Finally, notice that bounds from X-ray and $\gamma$-ray observations~\cite{Carr:2009jm, Ballesteros:2019exr, Arbey:2019vqx, Iguaz:2021irx, Chen:2021ngo, DeRocco:2019fjq, Laha:2019ssq, Dasgupta:2019cae, Laha:2020ivk, Coogan:2020tuf} and from BBN  abundances~\cite{Kohri:1999ex, Carr:2009jm, Acharya:2020jbv} and CMB anisotropies~\cite{Poulin:2016anj, Clark:2016nst, Stocker:2018avm, Poulter:2019ooo, Lucca:2019rxf, Acharya:2020jbv} are more stringent (see, e.g., Ref.~\cite{Carr:2020gox} for a complete compendium of bounds) than those obtained with neutrinos. All in all, these limits are complementary.

\subsection{Sensitivity for PBHs as DM}

\begin{figure}
	\centering
	\includegraphics[scale=0.7]{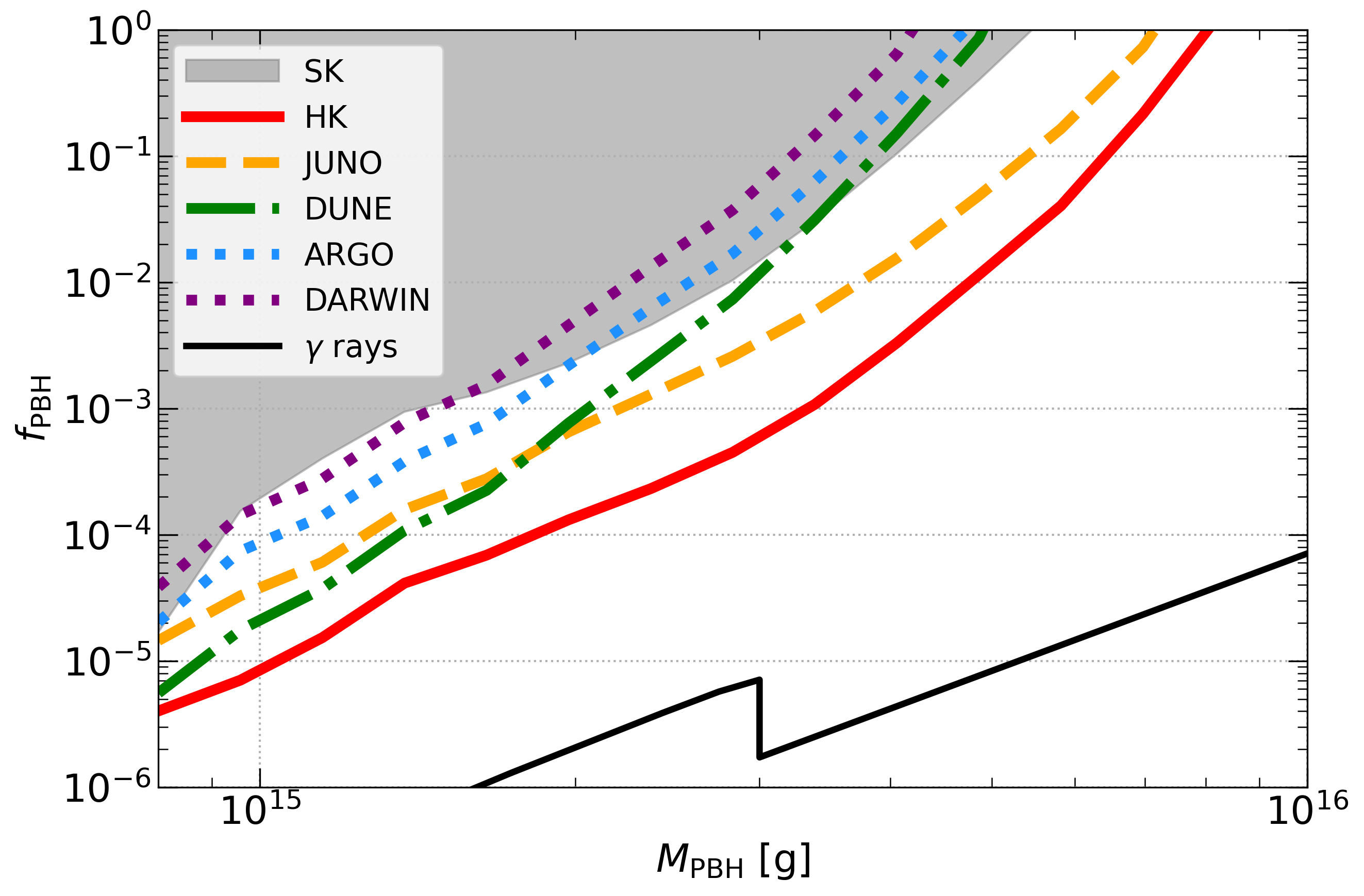}
	\caption{Current and projected (for 10~years of data-taking) 90\% CL limits on the fraction of PBHs as DM, $f_{\rm PBH}$, as a function of the PBH mass, $\Mbh \geq 8 \times 10^{14}$~g, for a monochromatic mass distribution and for the detectors discussed in this work: HK (red solid curve), JUNO (orange dashed curve), DUNE (green dot-dashed curve), DARWIN (purple dotted curve) and ARGO (blue dotted curve). Current limits from SK data are also indicated by the shaded region and bounds from $\gamma$-ray observations~\cite{Chen:2021ngo, Coogan:2020tuf} are indicated by the black solid line.}
	\label{fig:PBHbounds_DM}
\end{figure}

\begin{figure}
	\centering
	\includegraphics[scale=0.35]{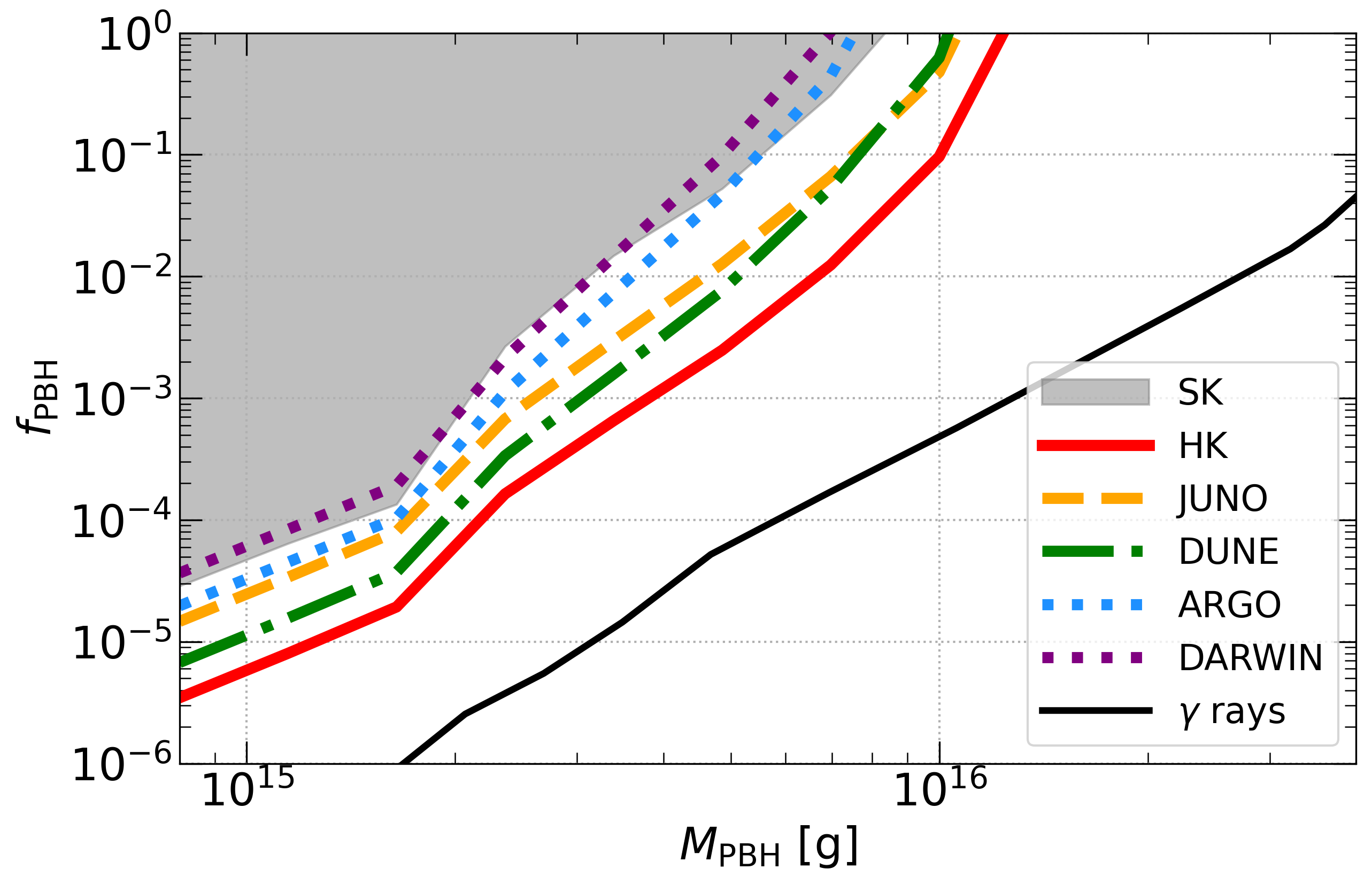}
	\includegraphics[scale=0.35]{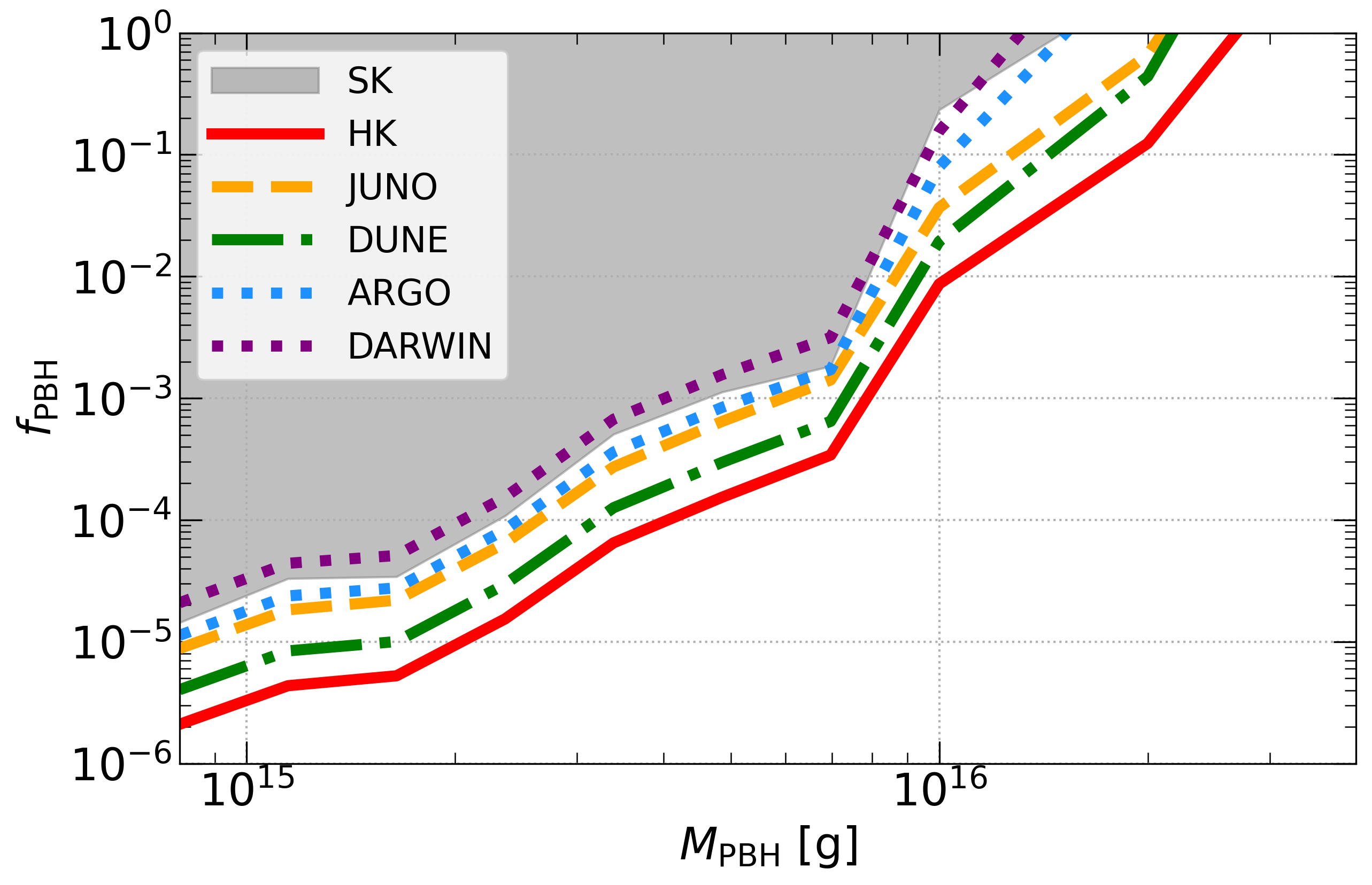}
	\caption{Same as Fig.~\ref{fig:PBHbounds_DM}, but for log-normal mass distributions, for two values of the width $\sigma = 0.5$ (left panel) and $\sigma = 1$ (right panel). Results are shown as a function of the median PBH mass. Current bounds from  $\gamma$-ray observations of the isotropic extragalactic background~\cite{Arbey:2019vqx} are shown for $\sigma=0.5$. For $\sigma=1$ the most stringent bounds come from CMB anisotropies~\cite{Acharya:2020jbv} and lie below the range shown in the plots.}
	\label{fig:PBHbounds_lognormal_DM}
\end{figure}

For all detectors we consider, we compute the $90\%$ CL ($\Delta \chi^2 = 2.71$) limits on the fraction of PBHs as DM, $f_{\rm PBH}$, as a function of the PBH mass, which are shown in Fig.~\ref{fig:PBHbounds_DM} for a monochromatic mass distribution and in Fig.~\ref{fig:PBHbounds_lognormal_DM} for log-normal mass distributions. Current bounds from $\gamma$-ray observations~\cite{Chen:2021ngo, Coogan:2020tuf} are also depicted in Fig.~\ref{fig:PBHbounds_DM}. In the case of a log-normal mass distribution with $\sigma = 1$, the most stringent bounds in the mass range we consider come from CMB anisotropies~\cite{Acharya:2020jbv} and lie in the interval $f_{\rm PBH} \sim 10^{-9} - 10^{-6}$ (i.e., about three or four orders of magnitude more restrictive than future expectations for HK). For a log-normal mass distribution with $\sigma=0.5$, CMB limits lie between those of the monochromatic and the log-normal distribution with $\sigma=1$~\cite{Acharya:2020jbv} and are similar to those from $\gamma$-ray observations of the isotropic extragalactic background~\cite{Arbey:2019vqx}.

Clearly, HK will outperform the other detectors we consider, mainly due to its much larger mass. Nonetheless, different methods of detection and backgrounds result in distinct bounds, even with similar exposures. It is the case of JUNO, which, for a similar exposure, would improve the current bound from SK by a factor of a few. Likewise, it is important to remark that the relative sensitivity of each detector also depends on the PBH mass. For instance, for a monochromatic mass distribution, DUNE will be able to obtain more stringent bounds than JUNO for $M_{\rm PBH} \lesssim 4 \times 10^{15}$~g, while for $M_{\rm PBH} \gtrsim 4 \times 10^{15}$~g the bounds from JUNO are expected to be more restrictive. These differences can be understood from the signal and background event spectra, which differ for each channel and detector: JUNO becomes more sensitive than DUNE to spectra peaked at lower energies, which correspond to larger PBH masses, due to its better neutrino energy resolution. On another hand, although the expected limits from DM detectors are weaker due to their smaller size, a detector $\sim 60$ times smaller than SK as ARGO could reach limits comparable to (and even slightly better than) current bounds from SK data. Finally, as shown in Fig.~\ref{fig:PBHbounds_lognormal_DM}, for log-normal mass distributions, the limits are more restrictive (except for $\Mbh \lesssim 10^{15}$~g) and extend to more massive PBHs the broader the mass distribution, in analogy to limits from the isotropic extragalactic $\gamma$-ray background~\cite{Arbey:2019vqx}. Furthermore, a wider mass spectrum implies that a wider detection range is sensitive to the neutrino flux, so that JUNO becomes less competitive than DUNE up to larger masses.

In all these analyses, the dominant backgrounds above the chosen energy threshold are produced by solar and atmospheric neutrinos. Nevertheless, the DSNB flux is a guaranteed, yet uncertain, background for the signal from PBH evaporation. Therefore, we have also checked the effect of adding this extra background component by recomputing all the bounds. Within current limits on the DSNB flux~\cite{Super-Kamiokande:2021jaq}, its addition would worsen future limits on the PBH abundance by less than a factor of two in the case of neutrino detectors, and by much less ($\lesssim 20\%$) in the case of DM detectors, in the entire mass range we consider. The differences are more important for the most massive non-evaporated PBHs. Yet, given the small impact of the (uncertain) DSNB flux on the results, we do not show explicitly these limits, which are quite similar to the ones we depict, that are slightly more optimistic.

As mentioned above, other studies in the literature have already presented forecasts for some of the detectors discussed here. We are in reasonably good agreement with all of them, although some differences arise. Projected sensitivities for the JUNO detector, for the case of monochromatic mass distributions, were recently obtained~\cite{Wang:2020uvi}. The constraints are slightly more stringent than ours, which could be due to a couple of differing points in the analyses: in Ref.~\cite{Wang:2020uvi} the total integrated number of events within an energy range (different for each PBH mass being tested) was used, rather than accounting for the spectral information; and the neutrino fluxes are about a factor of two larger than ours (which are in agreement with those in Ref.~\cite{DeRomeri:2021xgy}) in the relevant energy range. On another hand, Ref.~\cite{DeRomeri:2021xgy} estimated the sensitivity of the DUNE detector to neutrinos from non-evaporated PBHs, for the same mass distributions considered in our work. There are several differences between our treatment and the one followed in Ref.~\cite{DeRomeri:2021xgy}, such as the characterization of the detector properties, in particular the energy resolution and efficiency; the statistical analysis, as we do not include systematic uncertainties on the background and we do not include the contribution from the DSNB either; and the energy threshold, which we take at a slightly higher value (19~MeV rather than 16~MeV). Yet, our results for non-evaporated PBHs are in good agreement with those in Ref.~\cite{DeRomeri:2021xgy}. Likewise, for a monochromatic mass distribution, forecast limits for liquid xenon detectors such as DARWIN were obtained in Ref.~\cite{Calabrese:2021zfq}, whose analysis also presents some differences with respect to ours: we include the energy resolution of the detectors and our energy range is wider; Ref.~\cite{Calabrese:2021zfq} includes systematic uncertainties on the normalization of backgrounds, and the DSNB is also included; and the neutrino fluxes seem to be larger than ours.\footnote{This could be due their adding the output of {\tt BlackHawk} for primary and secondary spectra (see footnote~\ref{fn:BHoutput}).} All in all, our results for 10~years with DARWIN are slightly less stringent, as they are very similar to those for 200~kton$\times$year in Ref.~\cite{Calabrese:2021zfq}.

\subsection{Sensitivity for evaporated PBHs}

\begin{figure}
	\centering
	\includegraphics[scale=0.7]{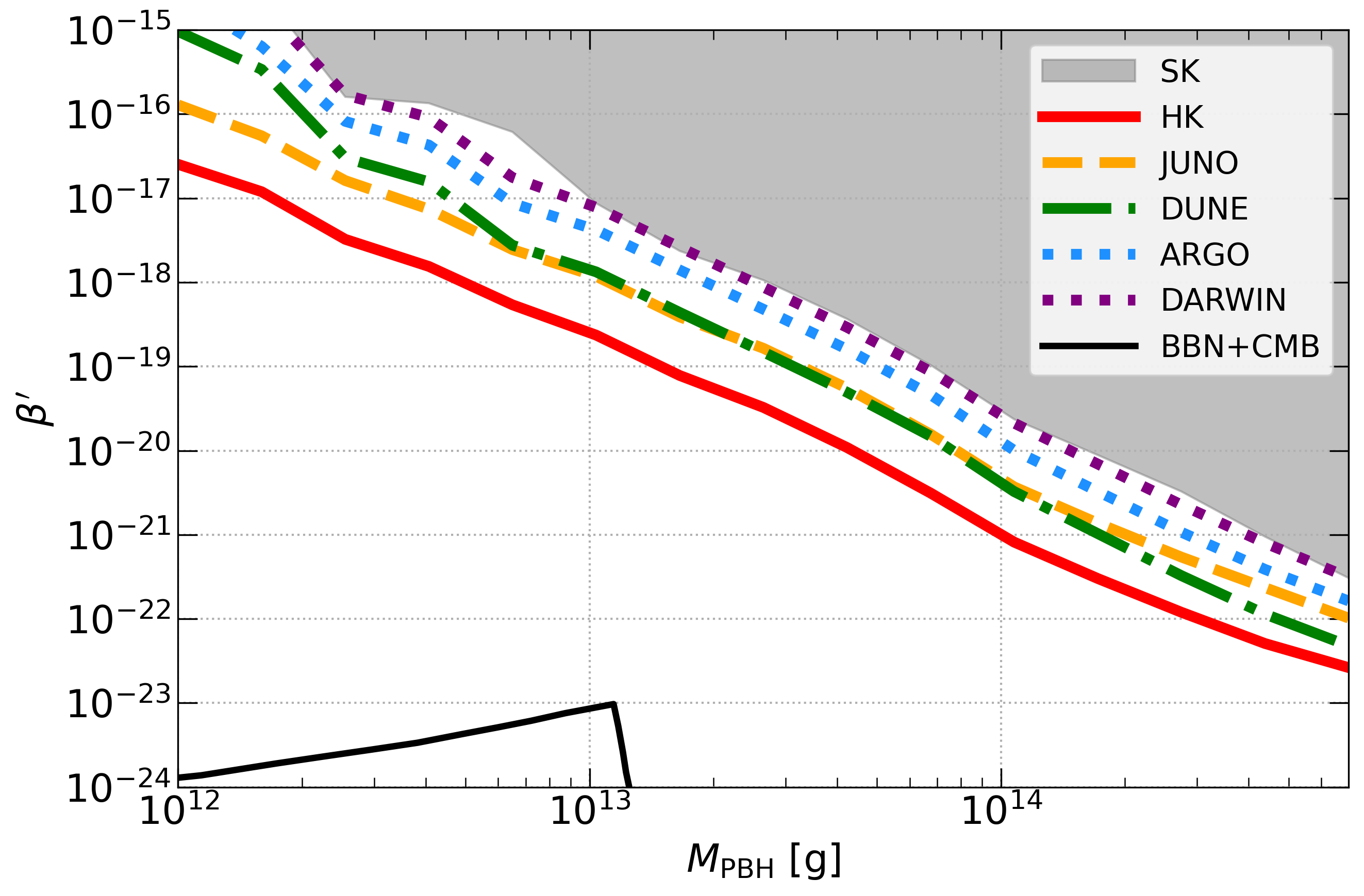}
	\caption{Same as Fig.~\ref{fig:PBHbounds_DM} but for evaporated PBHs, $\Mbh < 8 \times 10^{14}$~g. Current bounds from BBN and CMB~\cite{Acharya:2020jbv} are indicated by the black solid line. For $M_{\rm PBH} \gtrsim 3 \times 10^{14}$~g, the most important limits come from $\gamma$-ray observations of the isotropic extragalactic background~\cite{Chen:2021ngo}, but lie at $\beta' \lesssim 10^{-25}$.}
	\label{fig:PBHbounds_evap}\vspace{2cm}
\end{figure}

\begin{figure}[h]
	\centering
	\includegraphics[scale=0.35]{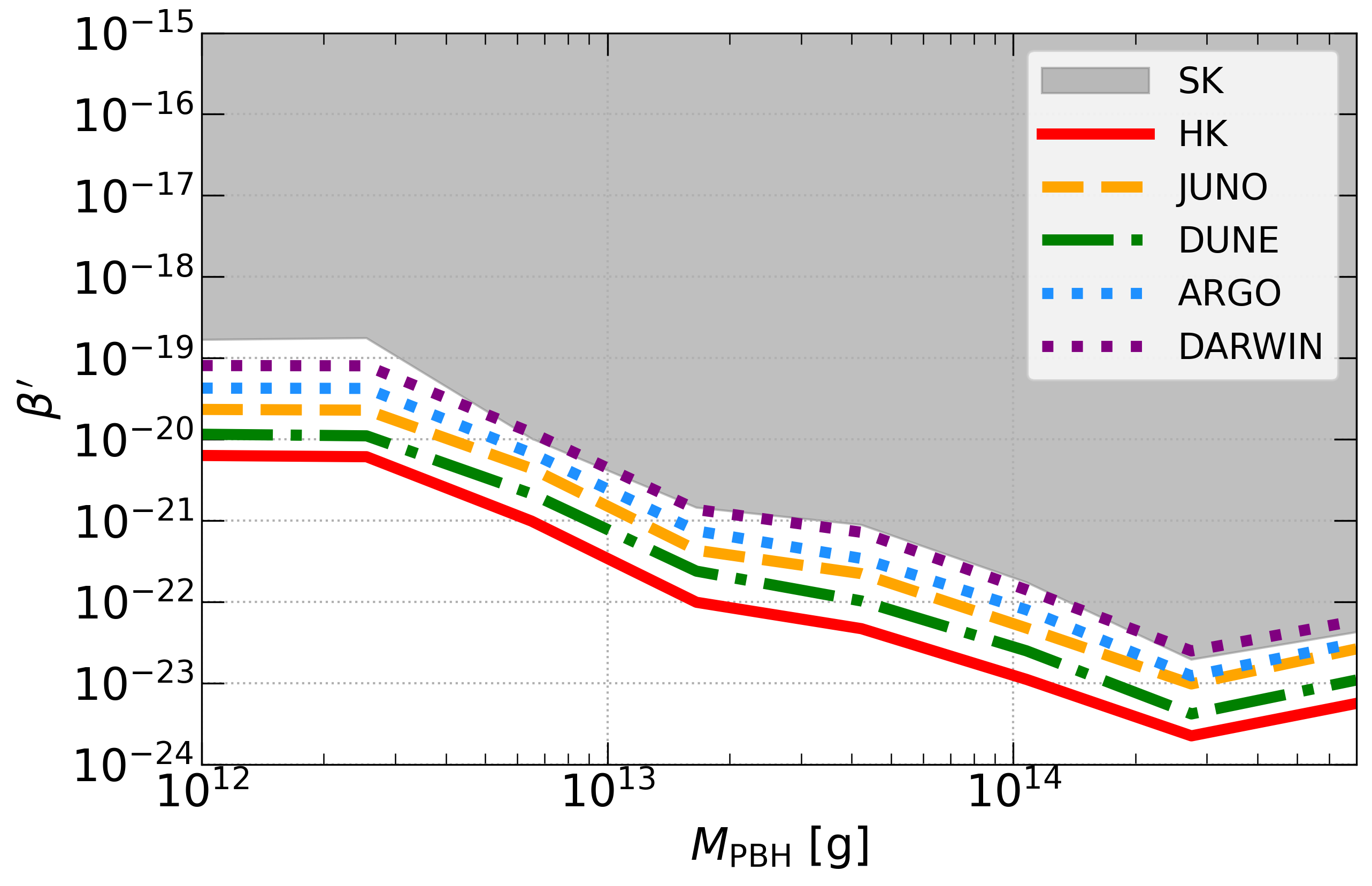}
	\includegraphics[scale=0.35]{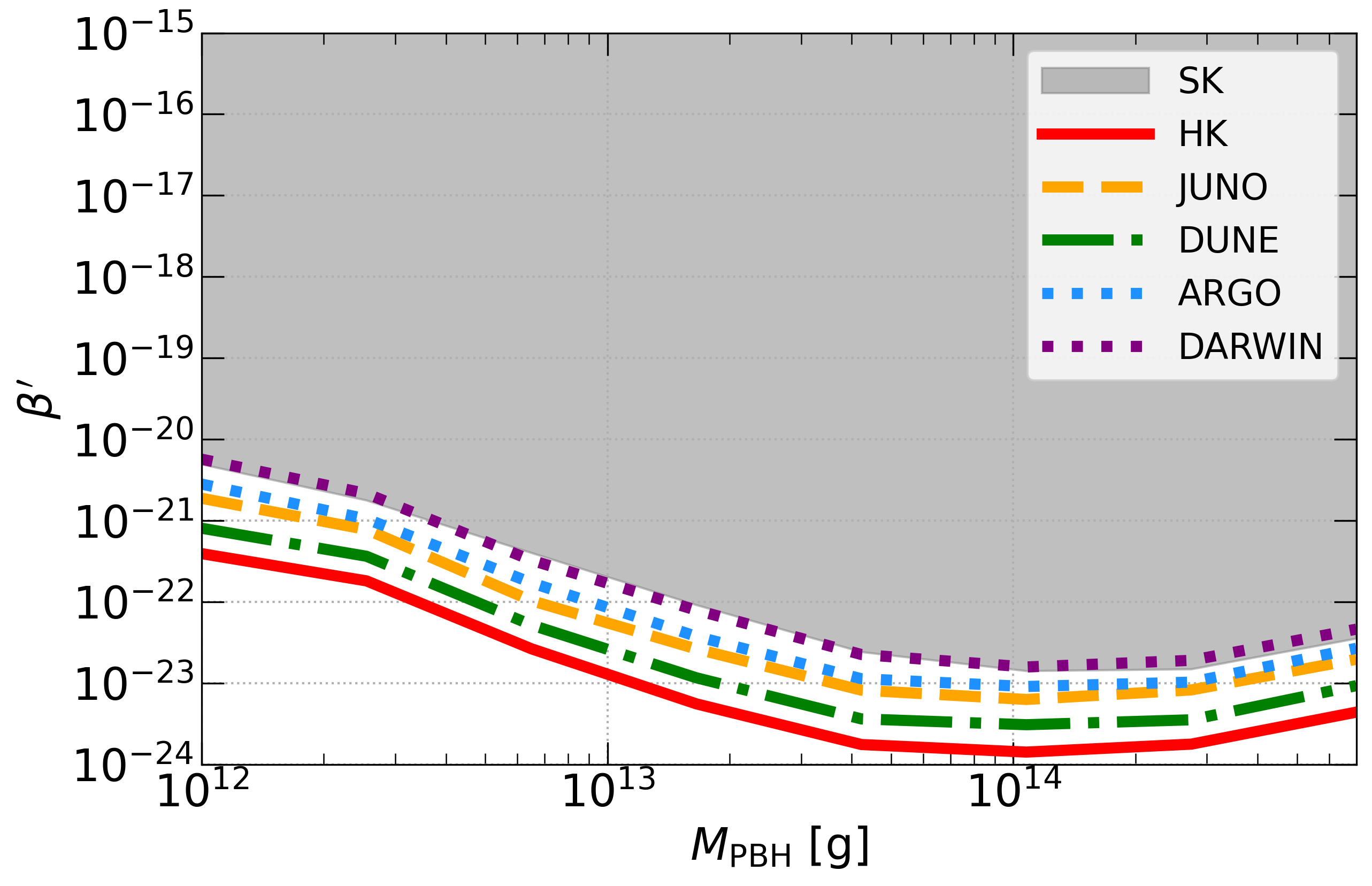}
	\caption{Same as Fig.~\ref{fig:PBHbounds_evap}, but for log-normal mass distributions, for two values of the width  $\sigma = 0.5$ (left panel) and $\sigma = 1$ (right panel). Results are shown as a function of the median PBH mass. The most stringent bounds come from CMB anisotropies~\cite{Acharya:2020jbv} and lie below the range shown in the plots.}
	\label{fig:PBHbounds_lognormal_evap}
\end{figure}

PBHs with masses below $\sim M_{\rm evap}$ would have already completely evaporated today, and thus cannot form part of the DM now. It is customary, thus, to constrain their abundance at any time via their initial abundance $\beta$, or the modified parameter $\beta'$, defined in Eq.~\eqref{eq:betaprime}. For monochromatic and log-normal mass distributions, $90\%$ CL ($\Delta \chi^2 = 2.71$) forecast limits on the initial abundance $\beta'$ versus PBH mass are depicted in Figs.~\ref{fig:PBHbounds_evap} and~\ref{fig:PBHbounds_lognormal_evap}, respectively. In Fig.~\ref{fig:PBHbounds_evap}, we also show the most stringent bounds for the monochromatic case, which are obtained from BBN abundances and CMB anisotropies~\cite{Acharya:2020jbv}. For $M_{\rm PBH} \gtrsim 3 \times 10^{14}$~g, the most stringent limits come from $\gamma$-ray observations of the isotropic extragalactic background~\cite{Chen:2021ngo}, although they are below the range shown in the figures, reaching $\beta' \sim 3 \times 10^{-26}$ for $M_{\rm PBH} \sim M_{\rm evap}$. For log-normal mass distributions with $\sigma=0.5$ and $\sigma=1$, CMB limits are the most important ones and, for the mass range we consider, lie in the interval $\beta' \sim 10^{-25} - 10^{-27}$ for $\sigma=0.5$ and $\beta' \sim 10^{-28} - 10^{-27}$ for $\sigma=1$~\cite{Acharya:2020jbv}, which are several orders of magnitude more stringent than future expectations for HK.

One can see differences among detectors analogous to those illustrated in Figs.~\ref{fig:PBHbounds_DM} and~\ref{fig:PBHbounds_lognormal_evap}. Exposure is again the key property for constraining the PBH abundance, being HK the detector with the best sensitivity. Nevertheless, different channels also vary the mass dependence on the bounds. For instance, as discussed for the case of non-evaporated PBHs, JUNO is slightly more sensitive than DUNE to lower energies, so JUNO is slightly more constraining than DUNE for the smallest masses, whereas for $\Mbh \gtrsim 10^{14}$~g DUNE would be expected to obtain more restrictive limits.

In this PBH mass range, we have also checked the impact of adding the DSNB flux as another background component. The differences are similar to those obtained for non-evaporated PBHs, in this case being more important for the least massive PBHs, but the limits never worsen more than a factor of two in the entire mass range we consider. Therefore, given the small differences, as for non-evaporated PBHs, we do not include the (uncertain) DSNB flux on the results we show.

Finally, note that the only previous constraints on evaporated PBHs ($\Mbh < M_{\rm evap}$) using neutrino detectors were obtained in Ref.~\cite{Carr:2009jm} for monochromatic mass distributions and with early SK limits on the DSNB. Here, we improve over those results for a monochromatic mass distribution (see Sec.~\ref{sec:SK}) and also obtain results for extended mass distributions for several future neutrino detectors. The limits from these extended mass distributions can be several orders of magnitude more stringent than for the monochromatic case.

\section{Conclusions}
\label{sec:conclusions}

The existence of PBHs formed in the early Universe from large inhomogeneities is an exciting possibility. If they exist, during their evaporation, PBHs would unavoidably Hawking radiate all types of particles, and in particular, neutrinos. In this work, we have evaluated constraints on the abundance of comet-mass PBHs from current and future measurements of this neutrino flux. Specifically, we have considered non-rotating PBHs with monochromatic and extended (log-normal) mass distributions, covering a large PBH mass range, spanning from $10^{12}$~g to $\sim 10^{16}$~g.

Low mass PBHs (less massive than $\sim 8 \times 10^{14}$~g), with lifetimes shorter than the age of the Universe, would have already fully evaporated, leaving behind a relic emission of plausibly detectable neutrinos. On another hand, PBHs with lifetimes longer than the age of the Universe would be present today, contributing to a fraction of the total DM abundance. Both, evaporated and non-evaporated PBHs could be probed by their neutrino emission, and limits on either the initial or current abundance of such objects could be set (or a positive signal inferred) by comparing the neutrino-induced events in neutrino detectors with the expected backgrounds. Previous estimates with actual data considered the limit on the integrated DSNB flux obtained from SK data and set bounds on the maximum fraction of DM in the form of PBHs by imposing the integrated flux of neutrinos from PBHs (in a given energy range) to be smaller~\cite{Carr:2009jm, Dasgupta:2019cae}. Here, we have revisited and improved upon those analyses, including in detail the spectral information of the potential PBH signal and properly accounting for the backgrounds. Thus, we provide more detailed bounds on the PBH abundance from SK data. In addition to non-evaporating PBHs, we have also considered the potential neutrino flux from fully evaporated PBHs, and we have set limits on their initial fraction of the total energy density. The updated bounds from SK on the PBH abundance as a function of the mass for monochromatic and log-normal mass distributions, both in the case of fully evaporated and non-evaporated PBHs are depicted in Fig.~\ref{fig:PBHbounds_SK}.

Furthermore, several future neutrino experiments, such as HK, JUNO, and DUNE, have been shown to provide high sensitivities which would outperform current constraints from SK, using different detection channels and in some cases, with larger exposures. DM direct detection experiments able to reach the so-called neutrino floor, such as ARGO and DARWIN, are however unlikely to significantly improve the current bounds from SK, given their smaller relative masses. These current and future neutrino limits on the abundance of PBHs are depicted in Figs.~\ref{fig:PBHbounds_DM}--\ref{fig:PBHbounds_lognormal_evap}, for monochromatic and log-normal mass distributions. 

In this way, we have studied current and future neutrino limits on the abundance of non-rotating PBHs and have provided a general picture of these results by comparing the limits and sensitivies of different types of detectors. Finally note, however, that both, current and projected constraints, are expected to be weaker than current limits from X-ray and $\gamma$-ray observations~\cite{Carr:2009jm, Ballesteros:2019exr, Arbey:2019vqx, Iguaz:2021irx, Chen:2021ngo, DeRocco:2019fjq, Laha:2019ssq, Dasgupta:2019cae, Laha:2020ivk, Coogan:2020tuf} (see, e.g., Refs.~\cite{Coogan:2020tuf, Ray:2021mxu} for forecasts of future limits), from the primordial abundance of light elements~\cite{Kohri:1999ex, Carr:2009jm, Acharya:2020jbv} and from CMB anisotropies~\cite{Poulin:2016anj, Clark:2016nst, Stocker:2018avm, Poulter:2019ooo, Lucca:2019rxf, Acharya:2020jbv}, either for PBHs already evaporated or for PBHs constituting a fraction of DM. Nevertheless, we stress that the present study provides an independent and complementary way of testing the abundance of PBHs.

\section*{Data availability}

The code employed in this article to compute the neutrino events and the forecasts for future experiments, {\tt nuHawkHunter} \cite{pablo_villanueva_domingo_2022_6380867}, is publicly available on~\href{https://github.com/vmmunoza/nuHawkHunter}{GitHub \faGithub}.%
\footnote{\url{https://github.com/vmmunoza/nuHawkHunter}}

\section*{Acknowledgments}

We thank Jérémy Auffinger for his help with the {\tt BlackHawk} code. We thank Peter Denton, Valentina de Romeri, Pablo Martínez-Miravé, Joannis Papavassiliou, Anna Suliga and Mariam Tórtola for enlightening discussions.
NB received funding from the Patrimonio Autónomo - Fondo Nacional de Financiamiento para la Ciencia, la Tecnología y la Innovación Francisco José de Caldas (MinCiencias - Colombia) grants 80740-465-2020 and 80740-492-2021.
NB and SPR are supported by the Spanish MCIN/AEI/10.13039/501100011033 grant PID2020-113334GB-I00.
The work of VM is supported by CONICYT PFCHA/DOC\-TO\-RA\-DO BECAS CHILE/2018 - 72180000. 
SPR is also partially supported by the Portuguese FCT (CERN/FIS-PAR/0004/2019). 
PVD acknowledges support from the Generalitat Valenciana through the GenT program (CIDEGENT/2018/019, CPI-21-108). This project has also received funding/support from the European Union's Horizon 2020 research and innovation program under the Marie Skłodowska-Curie grant agreement No 860881-HIDDeN.

\bibliographystyle{JHEP}
\bibliography{biblio}

\end{document}